\documentclass[twocolumn,times]{aastex63}
\usepackage[utf8]{inputenc}
\usepackage[normalem]{ulem}
\usepackage{amsmath}

\begin{document}

\title{An In Situ Study of Turbulence Near Stellar Bow Shocks}
\author[0000-0002-4941-5333]{Stella Koch Ocker}
\author[0000-0002-4049-1882]{James M. Cordes}
\author[0000-0002-2878-1502]{Shami Chatterjee}
\affiliation{Department of Astronomy and Cornell Center for Astrophysics and Planetary Science, Cornell University, Ithaca, NY, 14853, USA}
\author[0000-0001-8885-6388]{Timothy Dolch}
\affiliation{Department of Physics, Hillsdale College, 33 E. College Street, Hillsdale, MI 49242, USA}
\affiliation{Eureka Scientific, Inc.  2452 Delmer Street, Suite 100, Oakland, CA 94602-3017}
\correspondingauthor{Stella Koch Ocker}
\email{sko36@cornell.edu}

\turnoffeditone

\begin{abstract}

Stellar bow shocks are observed in a variety of interstellar environments and are shaped by the conditions of gas in the interstellar medium (ISM). In situ measurements of turbulent density fluctuations near stellar bow shocks are only achievable with a few observational probes, including H$\alpha$ emitting bow shocks and the \textit{Voyager Interstellar Mission (VIM)}. In this paper, we examine density variations around the Guitar Nebula, an H$\alpha$ bow shock associated with PSR B2224$+$65, in tandem with density variations probed by \textit{VIM} near the boundary of the solar wind and ISM. High-resolution \textit{Hubble Space Telescope} observations of the Guitar Nebula taken between 1994 and 2006 trace density variations over scales from 100s to 1000s of au, while \textit{VIM} density measurements made with the Voyager 1 Plasma Wave System constrain variations from 1000s of meters to 10s of au. The power spectrum of density fluctuations constrains the amplitude of the turbulence wavenumber spectrum near the Guitar Nebula to ${\rm log}_{10}C_{\rm n}^2 = -0.8\pm0.2$ m$^{-20/3}$ and for the very local ISM probed by Voyager ${\rm log}_{10}C_{\rm n}^2 = -1.57\pm0.02$ m$^{-20/3}$. Spectral amplitudes obtained from multi-epoch observations of four other H$\alpha$ bow shocks also show significant enhancements from values that are considered typical for the diffuse, warm ionized medium, suggesting that density fluctuations near these bow shocks may be amplified by shock interactions with the surrounding medium, or by selection effects that favor H$\alpha$ emission from bow shocks embedded in denser media. 

\end{abstract}

\section{Introduction}
Bow shocks form around stars moving at supersonic and super-Alfv\'enic speeds through the interstellar medium (ISM), and their morphologies are determined by the conditions for \edit1{ram} pressure balance between their stellar winds and the surrounding interstellar gas. Bow shocks are observed around stars at a range of life stages, including runaway OB stars \citep{2012A&A...538A.108P, 2015A&A...578A..45P} and supergiants \citep{2012A&A...548A.113D}, and their signatures are observed over wavelengths spanning radio to X-rays. Neutron stars, in particular, are believed to generally produce bow shocks as they are born at speeds ranging from 100s to 1000s of kilometers per second and typically move faster than the speed of sound once they have exited their supernova remnants and entered the ambient ISM. Currently, one of the only direct methods for detecting neutron star bow shocks is by observing the H$\alpha$ emission that is produced by collisional excitation of interstellar gas at the bow shock, but this method requires that the gas be partially neutral, and H$\alpha$ bow shocks have only been observed from about 9 neutron stars thus far \citep{2014ApJ...784..154B}. Observations of nonthermal radio and X-ray emission from ram pressure confined pulsar wind nebulae (PWNe) have provided indirect evidence for the presence of bow shocks around an additional handful of neutron stars  \citep[][]{2017JPlPh..83e6301K}, but unlike H$\alpha$-emitting bow shocks, these observations do not yield direct measurements of the bow shock stand-off radius. Far-ultraviolet emission from pulsar bow shocks may also yield estimates of the stand-off radius, but has only been detected from two pulsars thus far \citep{2016ApJ...831..129R, 2017ApJ...835..264R}. Since the stand-off radius is directly related to the interstellar gas density, H$\alpha$ measurements of the bow shock stand-off radius over time are thus one of the only direct probes of turbulent density fluctuations in the partially ionized ISM. 

\indent This method is perhaps best exemplified by the Guitar Nebula (GN), the unusually shaped bow shock formed around the radio pulsar B2224$+$65. The GN was initially detected with the 5 meter Palomar telescope in H$\alpha$ emission extending over about an arcminute on the sky \citep{1993Natur.362..133C}, and follow-up observations of the bow shock nose with the \textit{Hubble Space Telescope (HST)} in 1994 and 2001 were able to resolve changes in the stand-off radius and interstellar density over the seven year timescale \citep{2002ApJ...575..407C, 2004ApJ...600L..51C}. Observations with the \textit{Discovery Channel Telescope} in 2014 demonstrated continued large-scale evolution of the entire nebula, consistent with an expansion rate of about 200 km/s \citep[][Dolch et al. in preparation]{2016JASS...33..167D}.  Magnetohydrodynamic modeling has confirmed that the bow shock's large-scale, quasi-oscillatory morphology can be predominantly ascribed to density variations in the surrounding medium \citep{2017MNRAS.464.3297Y,2019MNRAS.484.1475T,2020MNRAS.497.2605B}. However, the detection of multiple glitches in the pulsar's spin period \citep{2006AA...457..611J, 2010MNRAS.404..289Y} also raises the question whether changes in the pulsar's spin-down luminosity can modify the observed bow shock, and how, if at all, the pulsar wind modifies density fluctuations in the surrounding ISM \citep{2002ApJ...575..407C, 2016JASS...33..167D}. 

\indent Pulsar bow shocks are not the only in situ probe of turbulent density fluctuations in the ISM. The \textit{Voyager Interstellar Mission (VIM)} spacecrafts Voyager 1 and Voyager 2 (V1/V2) both directly sample electron density fluctuations in the very local ISM (VLISM), a region of space beyond the heliopause that is the boundary at which the solar wind and interstellar plasma reach pressure balance. \edit1{The structure of the outer heliosphere and VLISM can broadly be divided into three main regions: the termination shock, where the solar wind slows to subsonic speeds; the heliopause (i.e., the contact discontinuity), which is taken to be the boundary of the heliosphere; and a bow shock that sweeps up and decelerates the interstellar wind \citep[e.g.][]{2016SSRv..200..475O}.} The exact nature of the heliosphere's bow shock, if it exists, is unclear.  Measurements of the Sun's velocity with respect to the VLISM made by the \textit{Interstellar Boundary Explorer (IBEX)} initially suggested that the velocity is below the fast magnetosonic speed \citep{2012Sci...336.1291M}, implying that under certain conditions (such as a strong magnetic field $\sim 4$ $\mu$G) no bow shock would be present \citep{2013ApJ...763...20Z}. However, more recent measurements have revised the interstellar flow velocity to about 26 km s$^{-1}$ \citep{2015ApJ...801...28M, 2018ApJ...854..119S}, above the nominal fast magnetosonic speed. Moreover, different configurations of the interstellar flow velocity, magnetic field, and densities of ionized and neutral hydrogen can give rise to heliospheric bow shocks or bow waves exhibiting a wide range of properties \citep{2013ApJ...763...20Z, 2013GeoRL..40.2923Z}. Recent study of magnetic turbulence with \textit{VIM} \edit1{suggests that VLISM turbulence follows a Kolmogorov spectrum with an outer scale of about 0.01 pc}, which is broadly consistent with theoretical predictions for the distance to the bow wave/shock \citep{2018ApJ...854...20B, 2020ApJ...904...66L}.

\indent Regardless of the heliospheric bow shock's unresolved nature, direct sampling of the ISM with both \textit{VIM} and the GN offers a unique opportunity to study turbulent plasma near the boundaries between stellar winds and their interstellar environments across a wide range of physical conditions. While Voyager can probe density fluctuations on scales as small as a kilometer, H$\alpha$ images of the GN reveal density fluctuations across the bow shock's entire historical trajectory up to $\sim 0.1$ pc. The heliosphere is moving at approximately 26 km/s relative to the local interstellar plasma flow, whereas the GN is generated by a pulsar moving at 770 km/s \citep{2019ApJ...875..100D}. The local interstellar environments in both cases are also potentially quite different. The heliosphere lies near the edge of a partially ionized cloud within the Local Bubble \citep{2019ApJ...886...41L}. The GN, by contrast, is 831 pc away and lies about $6^\circ$ above the Galactic plane, in an extended region of warm, partially ionized gas that exhibits complicated filamentary structure \citep{2002ApJ...575..407C}.

\indent In this paper, we examine direct measurements of electron density fluctuations near stellar bow shocks using H$\alpha$ images of the GN and data taken by the V1 Plasma Wave System (PWS) instrument. In Section~\ref{sec:theory} we outline how the wavenumber spectrum of turbulent density fluctuations can be constrained by both in situ and integrated electron density measurements. A description of the \textit{HST} observations of the Guitar is provided in Section~\ref{sec:GNobs}; our analysis of the Guitar includes a third \textit{HST} epoch from 2006. Observed variations in the bow shock morphology are analyzed in Section~\ref{sec:GNanalysis}, and a new parallax distance for B2224$+$65 obtained through high-precision Very Long Baseline Interferometry \citep[VLBI;][]{2019ApJ...875..100D} allows us to place more precise constraints on the bow shock stand-off radius. In Section~\ref{sec:glitches} we consider the impact of pulsar glitches on the bow shock and find that even the largest glitch observed from B2224$+$65 would have had negligible impact on the observed bow shock structure. In Section~\ref{sec:VLISM} we discuss high-resolution density measurements obtained by V1 PWS. Finally, in Section~\ref{sec:spectrum}, constraints on the density wavenumber spectrum from the Guitar and V1 are discussed in the context of density fluctuations observed along pulsar lines-of-sight (LOS) throughout the local ISM, including four other neutron star bow shocks. Conclusions and remarks on future work are provided in Section~\ref{sec:conc}. 

\section{Density Wavenumber Spectrum}\label{sec:theory}

The wavenumber spectrum of electron density ($n_e$) fluctuations is modeled as a power-law of the form
\begin{equation}
    P_{\delta n_e} = C_{\rm n}^2 \mathbf{q}^{-\beta}, \hspace{0.1in} q_{\rm o} \leq q \leq q_{\rm i}
\end{equation}
where $\beta = 11/3$ for Kolmogorov turbulence, $C_{\rm n}^2$ is the spectral amplitude, and $\mathbf{q}$ is the wavenumber, which is related to the length scale $L$ by $q = 2\pi/L$. The spectrum extends from the outer scale $q_{\rm o} = 2\pi/l_{\rm o}$ to the inner scale $q_{\rm i} = 2\pi/l_{\rm i}$.

\subsection{In Situ Density Measurements}
In situ density measurements at two epochs ($n_{e,1},n_{e,2}$) correspond to a spatial offset $\delta x = v(t_2-t_1)$, where $v$ is the velocity of the spacecraft or the pulsar. A pairwise estimate of the density structure function is then 
\begin{equation}
    \widehat{D}_{n_e} = \langle [n_e(\mathbf{x}) - n_e(\mathbf{x}+\delta \mathbf{x})]^2\rangle \approx (n_{e1} - n_{e2})^2.
\end{equation}
\edit1{The model density structure function $D_{n_e}$ is related to the autocorrelation function (ACF) $R_{n_e}$ as $D_{n_e}(\delta \mathbf{x}) = 2[R_{n_e}(0) - R_{n_e}(\delta \mathbf{x})]$, where the ACF is given by 
\begin{equation}
    R_{n_e}(\delta \mathbf{x}) = \int d\mathbf{q}e^{i(\mathbf{q}\cdot\delta\mathbf{x})} P_{\delta n_e}(\mathbf{q}).
\end{equation}
}
Integrating over the 3D wavenumber spectrum thus yields an analytic relationship between the density structure function and the spectral amplitude:
\edit1{
\begin{equation}
    D_{n_e}(\delta\mathbf{x}) = 2\int d\mathbf{q}\big(1 - e^{i(\mathbf{q}\cdot\delta\mathbf{x})}\big)P_{\delta n_e}(\mathbf{q}).
\end{equation}
For an isotropic wavenumber spectrum and $q_{\rm o}\gg 2\pi/\delta x \gg q_{\rm i}$}, $D_{n_e} = K_{n_e}(\beta)C_{\rm n}^2(\delta x)^{\beta-3}$, where $K_{n_e}(\beta) = 4\pi\Gamma(\beta/2 - 1)\Gamma(4-\beta){\rm cos}(\pi(\beta-3)/2)/(\beta-3)\Gamma(\beta/2) \approx 20$ for $\beta = 11/3$ and $K_{n_e} = 2\pi^2$ for $\beta = 4$.
Individual point estimates of the spectral amplitude \edit1{$\widehat{C}_{\rm n}^2$} and wavenumber \edit1{$\widehat{q}$} can then be obtained as
\begin{equation}\label{eq:Cn2ne}
    \widehat{C}_{\rm n}^2 \approx \frac{\widehat{D}_{n_e}(\delta\mathbf{x})}{K_{n_e}(\beta)(\delta x)^{\beta-3}}, \hspace{0.1in} \widehat{q}\approx 2\pi/\delta x.
\end{equation}
A distribution of point estimates $\widehat{C}_{\rm n}^2$ can then be fit for a characteristic value, which we denote as $C_{\rm n}^2$. In the analysis that follows we adopt the Kolmogorov spectral index $\beta = 11/3$ as a fiducial value for directly estimating $\widehat{C}_{\rm n}^2$ from density measurements, \edit1{based on substantial evidence that turbulence in both the ambient ionized ISM and the VLISM is consistent with this assumption \citep{1995ApJ...443..209A, 2010ApJ...710..853C, 2018ApJ...854...20B, 2020ApJ...904...66L}}.

\subsection{Integrated Density Measurements}
Pulsar timing observations yield measurements of the integrated electron density or dispersion measure ${\rm DM} = \int_0^D n_e dl$ along the LOS to a pulsar at a distance $D$. Prolonged timing campaigns by pulsar timing arrays (PTAs) sample the DMs of $\sim 80$ pulsars about once a month over months to decades-long timespans \citep[e.g.,][]{2021ApJS..252....4A}. The observed DM variations over time DM$(t)$ can be used to trace stochastic density fluctuations along a pulsar LOS and hence constrain the wavenumber spectrum, but doing so requires correcting for deterministic DM variations that arise from the pulsar LOS crossing discrete structures in the ISM \citep{2016ApJ...821...66L, 2017ApJ...841..125J}. In the absence of these deterministic contributions to DM$(t)$, the DM structure function $D_{\rm DM}(\tau) = \langle [{\rm DM}(t+\tau) - {\rm DM}(t)]^2\rangle$ is directly related to the rms of the DM variations, which can be related to the amplitude of the density wavenumber spectrum as:
\begin{equation}\label{eq:DMCn2}
    \widehat{C}_{\rm n}^2 = \frac{1}{K_{\rm DM}}\bigg[\frac{D_{\rm DM}(\tau)}{D(v_{\rm eff,\perp}\tau)^{\beta-2}}\bigg], \hspace{0.1in} \widehat{q} \approx 2\pi/(v_{\rm eff,\perp} \tau)
\end{equation}
where $\tau$ is the time lag between two point estimates of DM, $K_{\rm DM} \approx 88.3$ for $\beta = 11/3$, \edit1{$D$ is the distance to the pulsar}, and $v_{\rm eff,\perp}$ is the pulsar's effective transverse velocity, which is related to the transverse velocities of the pulsar, observer, and interstellar phase screen \citep{2016ApJ...821...66L}. For a Kolmogorov process, the DM structure function has the form $D_{\rm DM}(\tau) \sim \tau^{5/3}$.

\section{Observations of the Guitar Nebula}\label{sec:GNobs}

\begin{figure*}
    \centering
    \includegraphics[width=0.9\textwidth]{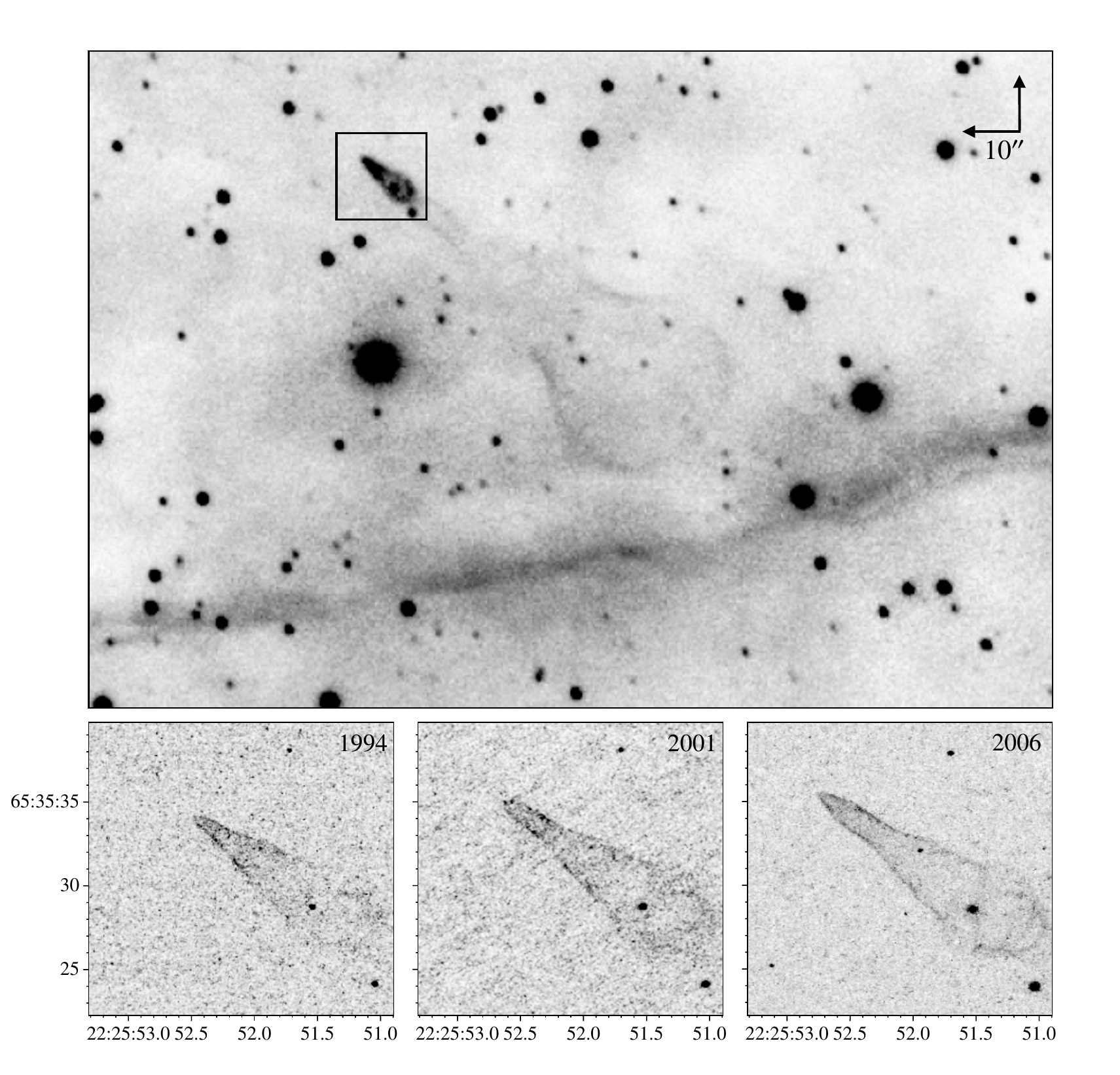}
    \caption{\textbf{Top:} H$\alpha$ image of the Guitar Nebula observed with the 5 m Hale Telescope at Palomar Observatory in 1995, previously discussed in \cite{2002ApJ...575..407C, 2004ApJ...600L..51C}. The black inset indicates the region shown in the bottom panels. The compass indicates north and east. \textbf{Bottom:} H$\alpha$ images of the head of the Guitar Nebula observed by \textit{HST} in 1994, 2001, and 2006. The 1994 and 2001 images were taken by WFPC2 and the 2006 image was taken by ACS. The 1994 and 2001 epochs were previously discussed in \cite{2002ApJ...575..407C, 2004ApJ...600L..51C}.}
    \label{fig:Halpha}
\end{figure*}

High-resolution observations of the GN were obtained with \textit{HST} in 1994 December and 2001 December using the Wide Field Planetary Camera 2 (WFPC2). The details of these observations have already been discussed in \cite{2002ApJ...575..407C, 2004ApJ...600L..51C}. A third \textit{HST} epoch was obtained in 2006 using Advanced Camera for Surveys (ACS) \citep{2013AAS...22115404G}. Figure~\ref{fig:Halpha} shows the H$\alpha$ images from all three epochs, which were aligned using the three brightest reference stars in each frame, in addition to a large-scale image of the Guitar taken at Palomar Observatory in 1995 \citep{2002ApJ...575..407C}. Between 1994 and 2001 the tip of the bow shock moved $1.32^{\prime \prime}$, equivalent to 1097 au, and between 2001 and 2006 the bow shock moved $0.86^{\prime \prime}$, equivalent to 715 au. The motion of the bow shock nose is consistent with the pulsar proper motion (see Table~\ref{tab:psr}).

\indent The shape of the bow shock nose can be directly inferred from the H$\alpha$ images and used to constrain the bow shock stand-off radius. In the thin-shell limit, the radial shape of the bow shock can be expressed as \citep{1996ApJ...459L..31W}
\begin{equation}\label{eq:Wilkin}
    R(\theta) = R_0~{\rm csc}~\theta\sqrt{3(1-\theta~{\rm cot}~\theta)}
\end{equation}
where $R_0$ is the stand-off radius and $\theta$ represents the angle between the pulsar's velocity and a point $R(\theta)$ along the bow shock. The stand-off radius is dictated by pressure balance between the ambient ISM and the neutron star wind, \edit1{and is directly related to the interstellar density $\rho_A$, the pulsar wind velocity $v_w$ and mass-loss rate $\dot{m}_w$, and the pulsar velocity $v_p$ \citep{1996ApJ...459L..31W}:
\begin{equation}\label{eq:R0}
    R_0 = \bigg(\frac{\dot{m}_w v_w}{4\pi\rho_A v_p^2}\bigg)^{1/2}.
\end{equation}
\added{Equation~\ref{eq:R0} assumes an isotropic stellar wind, which is not generally true of a relativistic neutron star wind. When the pulsar rotation vector is skewed relative to the pulsar proper motion and magnetic field, the pulsar wind will be quasi-isotropic, and its exact behavior will depend on both the orientations of these vectors and the opening angle of the wind. For B2224$+$65 the light-cylinder radius is significantly smaller than the bow shock stand-off radius, suggesting that an isotropic wind is an adequate assumption without additional evidence for anisotropy (for a detailed discussion see \citealt{2002ApJ...575..407C}).} Assuming that the spin-down energy loss is entirely carried by the relativistic wind, $\dot{m}_w v_w = \dot{E}/c$. } The stand-off radius can be conveniently re-formulated as an angle given by
\begin{equation}\label{eq:theta0}
    \theta_0 = 56.3 \hspace{0.05in} {\rm mas} \bigg(\frac{{\rm sin}^2 i}{n_A^{1/2}}\bigg) \bigg(\frac{\dot{E}_{33}^{1/2}}{\mu_{100}D_{\rm kpc}^2}\bigg),
\end{equation}
where $i$ is the inclination angle, \edit1{$n_A = \rho_A/m_{\rm H}$} is the total number density of the interstellar hydrogen and helium mixture in atomic mass units per cm$^{3}$, $\dot{E}$ is the spin-down luminosity in erg s$^{-1}$, $D$ is the distance to the pulsar in kpc, and $\mu_{100}$ is the pulsar proper motion in 100 mas yr$^{-1}$ \citep{2002ApJ...575..407C}. The spin-down luminosity is $\dot{E} = 4\pi^2 I\dot{P}/P^3$, where $I\approx 10^{45}$ g cm$^2$. The period derivative $\dot{P}$ can be corrected for the Schklovskii effect using 
\begin{equation}
    \dot{P} = \dot{P}_{\rm obs} - 2.43 \times 10^{-21} P \mu_{\rm masy}^2 D_{\rm kpc}
\end{equation}
where $\mu_{\rm masy}$ is the proper motion in mas yr$^{-1}$ \citep{1970SvA....13..562S, 2014ApJ...784..154B}. This correction is negligible for B2224$+$65. The atomic number density $n_A$ is converted to an electron density $n_e$ assuming a cosmic abundance $\gamma_{\rm H} = 1.37$, where $n_A = n_{\rm H}\gamma_{\rm H}$ and $n_{\rm H} \approx n_e$. The stand-off angle that is inferred from the outer edge of the H$\alpha$ emission corresponds to a forward shock that lies slightly upstream of the contact discontinuity, at an angular distance $\theta_a \approx 1.3\theta_0$ \citep{1992ApJ...400..638A, 2002AA...393..629B}. The period, period derivative, and spin-down luminosity for B2224$+$65 are shown in Table~\ref{tab:psr}. 

\begin{deluxetable*}{l C C C C C C}\label{tab:psr}
\tablecaption{H$\alpha$-Emitting Neutron Star Bow Shocks with Multi-Epoch Observations}
\tablehead{\multicolumn{7}{c}{Neutron Star Properties} \\ \hline \colhead{PSR} & \colhead{J0437$-$4715} & \colhead{B0740$-$28} & \colhead{J1741$-$2054} & \colhead{J2030$+$4415} & \colhead{J2124$-$3358} & \colhead{B2224$+$65} }
\startdata
DM (pc cm$^{-3}$) & 2.645 & 73.73 & 4.7 & \nodata & 4.6 & 36.1 \\
$P$ (s)  & 5.8\times10^{-3} & 0.167 & 0.41 & 0.308 & 4.9\times10^{-3} & 0.68 \\
$\dot{P}$ ($10^{-20}$ s s$^{-1}$) & 5.73 & 1.68\times10^{6} & 1.7\times10^{6} & 6.5\times10^5 & 2.06 &  9.66\times10^{5} \\
$\pi$ (mas)  & 6.396(54) & \nodata & \nodata & \nodata & 3.1(1) & 1.20(19) \\
$D$ (pc) & 156 & 2070 & 300 & 750 & 323 & 831  \\
$\mu_{\alpha}$ (mas yr$^{-1}$) & 121.679(52) & -29 & -63 & 15(11) & -14.14(4) & 147.22(23) \\
$\mu_{\delta}$ (mas yr$^{-1}$) & -71.820(86) & 4 & -89 & 84(12) & -50.08(9) & 126.53(19) \\
$v_T$ (km s$^{-1}$) & 105 & 287 & 155 & 303 & 78 & 765 \\
$\dot{E}_{33}$ (erg s$^{-1}$) & 5.5 & 140 & 9.5 & 22 & 6.8 & 1.2 \\ \hline \multicolumn{7}{c}{Bow Shock Properties}\\ \hline
Epoch 1 & 1993^{\rm a} & 2001^{\rm b} & 2009^{\rm c} & 2011^{\rm d} & 2001^{\rm e} & 1994 \\
Epoch 2 & 2012^{\rm d} & 2013^{\rm d} & 2015^{\rm f} & 2015^{\rm g} & 2013^{\rm d} & 2001 \\
Epoch 3 & \nodata & \nodata & \nodata & \nodata &  2015^{\rm h} & 2006  \\
$\theta_{a,1}$ ($^{\prime\prime}$) & 9.0 & 1.3 & 1.5^{\dagger} & 1.1 & 2.6 & 0.077  \\
$\theta_{a,2}$ ($^{\prime\prime}$) & 9.3 & 1.4 & \nodata & 0.5^{\dagger\dagger} & 5.0^{\dagger\dagger\dagger} & 0.11 \\
$\theta_{a,3}$ ($^{\prime\prime}$) & \nodata & \nodata & \nodata & \nodata & 2.73 & 0.094  \\
\enddata
\tablecomments{Parallax ($\pi$), distance ($D$), and proper motion in right ascension (including cos$\delta$) and declination ($\mu_\alpha, \mu_\delta$) are from the following references: \cite{2008ApJ...685L..67D} for J0437$-$4715,  \cite{2016MNRAS.455.1751R} for J2124$-$3358, and \cite{2019ApJ...875..100D} for B2224$+$65. The distances for B0740$-$28 and J1741$-$2054 are based on NE2001 \citep{2002astro.ph..7156C}. J2030$+$4415 is a radio-quiet pulsar and the quoted properties are based on $\gamma$-ray pulsations \citep{2012ApJ...744..105P} and X-ray astrometry \citep{2020ApJ...896L...7D}. All other neutron star properties are retrieved from the Australia Telescope National Facility (ATNF) Pulsar Catalogue \citep{2005AJ....129.1993M} unless otherwise noted. 
The bow shock apex distances $\theta_a$ are from the following references: (a) \cite{1993Natur.364..603B}, (b) \cite{2002AA...389L...1J}, (c) \cite{2010ApJ...724..908R}, (d) \cite{2014ApJ...784..154B}, (e) \cite{2002ApJ...580L.137G},
(f) \cite{2016ApJ...825..151M}, (g) \cite{2020ApJ...896L...7D}, (h) \cite{2017ApJ...851...61R}. 
Apex distances for B2224$+$65 are from this work.\\
$^\dagger$ \cite{2010ApJ...724..908R} find an inclination angle $i=80^\circ$ for J1741$-$2054. No $\theta_a$ is listed for the second epoch because \cite{2016ApJ...825..151M} find no evidence of a change in the stand-off radius. \\
$^{\dagger\dagger}$ This value of $\theta_a$ for J2030$+$4415 is only a nominal estimate from \cite{2020ApJ...896L...7D}, but this bow shock's complex, closed bubble morphology suggests more rigorous fitting for the inclination angle is needed.
$^{\dagger\dagger\dagger}$\cite{2014ApJ...784..154B} multiply $\theta_a$ for J2124$-$3358 by a factor of two to account for possible projection effects, but more detailed modeling by \cite{2017ApJ...851...61R} find $\theta_a$ is broadly consistent with the earlier \cite{2002ApJ...580L.137G} result but with $i\approx120^\circ$. }
\end{deluxetable*}

\section{Turbulence Around the Guitar}\label{sec:GNanalysis}

\subsection{Variations in the Stand-off Radius}\label{sec:GNanalysis1}
\begin{figure}
    \centering
    \includegraphics[width=0.45\textwidth]{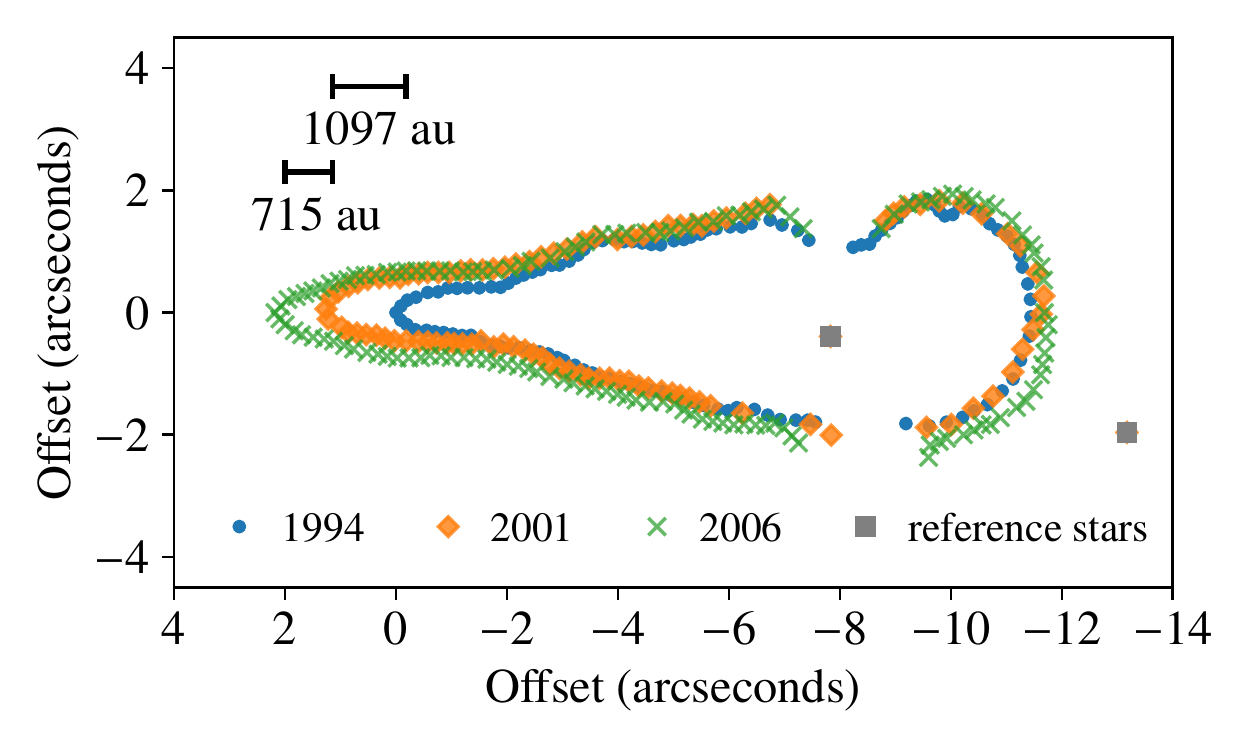}
    \caption{Outline of the limb-brightened head of the Guitar Nebula with the tip of the bow shock in 1994 set at the origin. The 1994, 2001, and 2006 epochs are shown as blue circles, orange diamonds, and green crosses, respectively. The grey squares show two of the three reference stars used to align the three epochs. The spatial offsets between the bow shock nose in 1994, 2001, and 2006 are also indicated.}
    \label{fig:GNoutline}
\end{figure}

Variations in the bow shock stand-off radius over time \edit1{for the GN} are \edit1{predominantly} related to density fluctuations in the surrounding gas over the length scales traversed by the pulsar between observations. \edit1{Changes in the pulsar wind velocity and mass-loss rate may also impact variations in the stand-off radius but are likely insignificant on the scales probed by the H$\alpha$ observations (see Section~\ref{sec:glitches})}. The outline of the bow shock nose was extracted from each H$\alpha$ image \edit1{by tracing local maxima in H$\alpha$ intensity along the edge of the shock} and is shown in Figure~\ref{fig:GNoutline}. Between 1994 and 2006 the bow shock nose moved $2.18^{\prime\prime}$, a total distance of 1812 au. While the overall morphology of the bow shock indicates a highly inhomogeneous interstellar density, the nose of the bow shock is adequately described by the thin-shell approximation given in Eq.~\ref{eq:Wilkin} to within about $2^{\prime\prime}$ of the nose. The projected angular apex distance $\theta_a$ was fit to \edit1{the bow shock outline extracted from} each \textit{HST} epoch \edit1{(see Figure~\ref{fig:GNoutline})} using least squares minimization of the $\chi^2$ statistic
\edit1{$\sum_{i=1}^{N} (M_i - D_i)^2/D_i^2$, where $M_i$ is the \cite{1996ApJ...459L..31W} model prediction for the location of a given point along the bow shock given by Eq.~\ref{eq:Wilkin} and $D_i$ is the corresponding location of a point along the observed shock.} \edit1{Given the extremely large transverse velocity of this pulsar, we do not consider the radial component of the pulsar velocity when fitting the apex shape.} The apex distance is $0.077(4)^{\prime\prime}$ in 1994, $0.11(1)^{\prime\prime}$ in 2001, and $0.094(6)^{\prime\prime}$ in 2006. Given negligible epoch-to-epoch changes in inclination angle, spin-down luminosity (see Section~\ref{sec:glitches}), proper motion, and distance, the ratio of the apparent stand-off angles between epochs gives the change in number density $\theta_{0,1}/\theta_{0,2} = \sqrt{n_{A,2}/n_{A,1}}$. We find $n_{A,2001}/n_{A,1994} = 0.5(1)$, $n_{A,2006}/n_{A,2001} = 1.4(3)$, and $n_{A,2006}/n_{A,1994} = 0.7(1)$. The first value is broadly consistent with the decrease in density found by \cite{2004ApJ...600L..51C}, and the additional density increase between 2001 and 2006 indicates that the bow shock's structure is influenced by quasi-oscillatory density variations on scales as small as 100s of au in the surrounding medium. 

\indent An unambiguous measurement of the electron density requires knowledge of the shock's inclination angle. Previous fitting by \cite{2002ApJ...575..407C, 2004ApJ...600L..51C} for the inclination angle marginally constrained the bow shock to lie in the plane of the sky, and the closed-off, ring-like morphology of the bow shock is broadly consistent with the \cite{2020MNRAS.497.2605B} MHD simulations of the shock for large inclination angles. Assuming $i = 90^\circ$ we find $n_{e,1994} = 0.44(5)$ cm$^{-3}$, $n_{e,2001} = 0.22(5)$ cm$^{-3}$, and $n_{e,2006} = 0.30(5)$ cm$^{-3}$. For an inclination angle $30^\circ$ smaller or larger, the densities will be about $56\%$ smaller. These densities are significantly larger than those inferred by \cite{2002ApJ...575..407C, 2004ApJ...600L..51C}, who used the NE2001 distance for the pulsar. The DM-derived distance based on NE2001 is 1.9 kpc, about twice as far as the recently observed parallax distance of 0.831 kpc \citep{2019ApJ...875..100D}, implying that the ISM along this LOS is denser than predicted by NE2001. 

\begin{figure*}
    \centering
    \includegraphics[width=\textwidth]{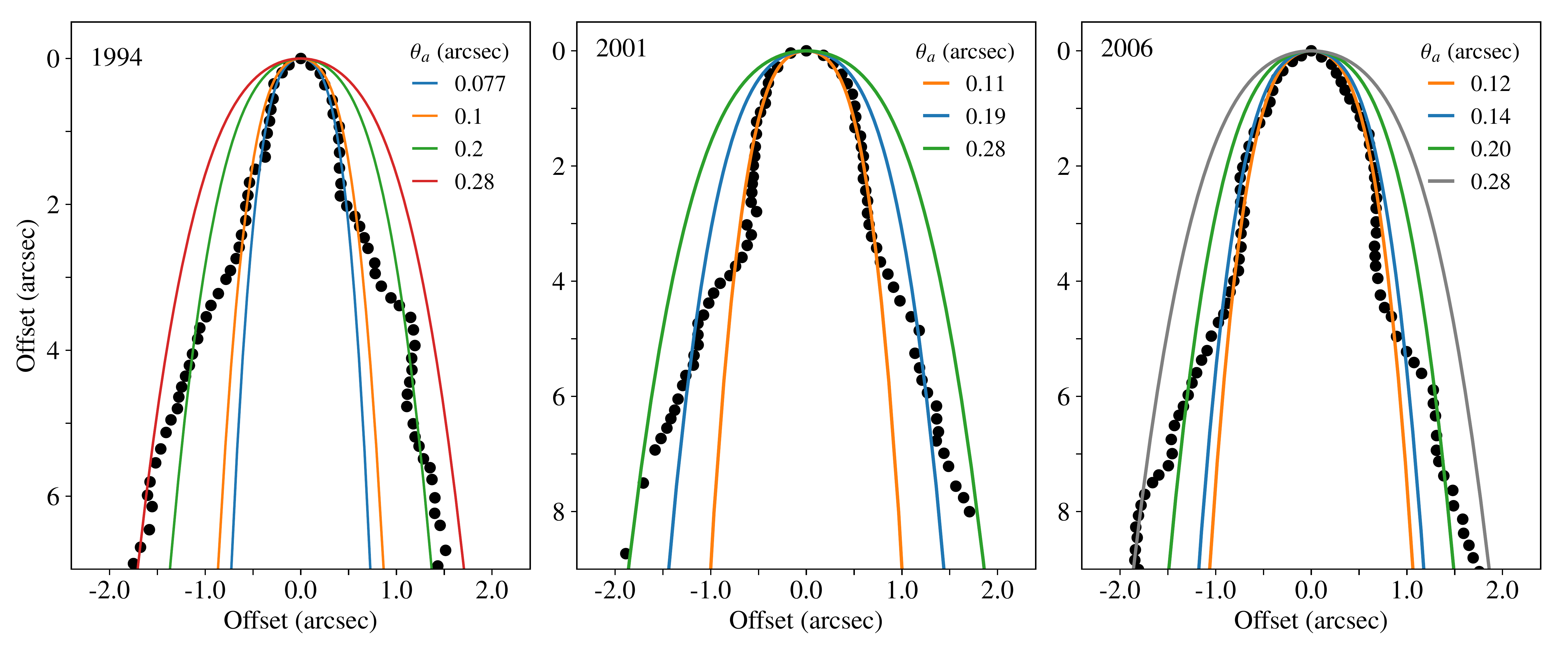}
    \caption{Outline of the limb-brightened head of the Guitar Nebula with the tip of the bow shock in each epoch set to the origin, which is aligned with the pulsar proper motion. The \cite{1996ApJ...459L..31W} model is shown for different values of the bow shock's projected angular apex distance. }
    \label{fig:GNbodypanel}
\end{figure*}

\indent While the \cite{1996ApJ...459L..31W} thin-shell model is only adequate near the tip of the Guitar, the morphology of the downstream shock can still be used to assess the scales on which the assumption of uniform density in the Wilkin model breaks down. Figure~\ref{fig:GNbodypanel} shows the outline of the head of the GN in each epoch with several realizations of the Wilkin model for different values of the projected stand-off radius. In each epoch, the downstream shock is significantly broader than predicted by the thin-shell model fit within $2^{\prime\prime}$ of the nose. The downstream shock is also consistent with density variations on roughly arcsecond scales, but these density variations do not have a consistent spatial periodicity. In addition, the bow shock structure is asymmetric, and this asymmetry does not necessarily translate uniformly from epoch to epoch. Variability in the observed asymmetry of the shock may be related to the shock's expansion into a medium that is nonuniform both parallel and transverse to the pulsar's proper motion, rather than related to, e.g., anisotropy in the pulsar wind \citep{2007MNRAS.374..793V}. However, Figure~\ref{fig:GNoutline} demonstrates that the downstream shock remains almost stationary between epochs, suggesting that the expansion rate downstream is significantly slower than at the nose. Apparent asymmetries in earlier observations of the shock may become smoothed out over time as thermal pressure modifies the morphology of the downstream shock, as was discussed in \cite{2007MNRAS.374..793V}. 

\indent \edit1{In Section~\ref{sec:spectrum} we calculate the turbulence spectrum of ionized interstellar gas around the GN using two methods:  1) epoch-to-epoch variations in the stand-off radius fit only near the shock apex, and 2) differences between stand-off radii fit to different portions of the downstream shock within each epoch. In the first case, epoch-to-epoch density variations are calculated using the stand-off radii fit to the shock apex (within $2^{\prime\prime}$) and an inclination angle $i = 90^\circ$ (see Eq.~\ref{eq:theta0}), and the resulting turbulence spectral amplitudes are calculated using Eq.~\ref{eq:Cn2ne}. The associated wavenumbers are calculated using the distance traversed by the pulsar between epochs. For three epochs (1994, 2001, and 2006), this epoch-to-epoch analysis yields three point estimates of the spectral amplitude and wavenumber. In the second method, we estimate density variations using stand-off radii fit to different portions of the downstream shock within a given epoch. For the models shown in Figure~\ref{fig:GNbodypanel}, we calculate the differences in stand-off radii between models that fit portions of the downstream shock and the model that fits within $2^{\prime\prime}$ of the apex. These differences yield rough estimates of density changes between gas around the downstream shock and gas near the shock apex, and hence an estimate of the turbulence spectral amplitude. This second method only considers inhomogeneity of the shock structure within individual epochs, rather than epoch-to-epoch changes. This second method also ignores a number of processes that could modify the morphology of the downstream shock (as discussed in the previous paragraph), and is less rigorous than epoch-to-epoch density variations inferred from only the shock apex. However, accounting for the inhomogeneity of the downstream shock vastly expands the range of wavenumbers that may be probed, from $10^{-15} \lesssim q \lesssim 10^{-13}$ m$^{-1}$. The error associated with fitting stand-off radii to the downstream shock is also orders of magnitude smaller than the spread of spectral amplitudes fit for the GN, which span over six orders of magnitude in the density fluctuation power spectrum. The turbulence spectral amplitudes obtained for the GN from both of the methods described are shown in Section~\ref{sec:spectrum}.}

\subsection{Glitches}\label{sec:glitches}
B2224$+$65 has five reported glitches between 1976 and 2007, the first and largest of which was reported by \cite{1982ApJ...255L..63B}. The glitch properties are shown in Table~\ref{tab:glitches}. The magnitude of a glitch is typically expressed as the fractional change in spin frequency $\Delta \nu/\nu$ and frequency derivative $\Delta \dot{\nu}/\dot{\nu}$ during a glitch. The fractional change in spin-down luminosity during a glitch, $\Delta \dot{E}/\dot{E}$, can be directly calculated from the change in $\nu$ and $\dot{\nu}$. For the simplest case where no change in frequency derivative is detected, $\Delta \dot{\nu} = 0$ and $\Delta \dot{E}/\dot{E} = \Delta \nu/\nu$. For typical cases where $|\Delta \dot{\nu}|>0$, $\Delta \dot{E}/\dot{E}$ must be explicitly calculated from the pre- and post-glitch timing solutions: $\Delta \dot{E}/\dot{E} = (\nu_2\dot{\nu}_2 - \nu_1\dot{\nu}_1)/\nu_1\dot{\nu}_1$, where subscripts $1$ and $2$ refer to pre- and post-glitch, respectively.

\begin{figure}
    \centering
    \includegraphics[width=0.48\textwidth]{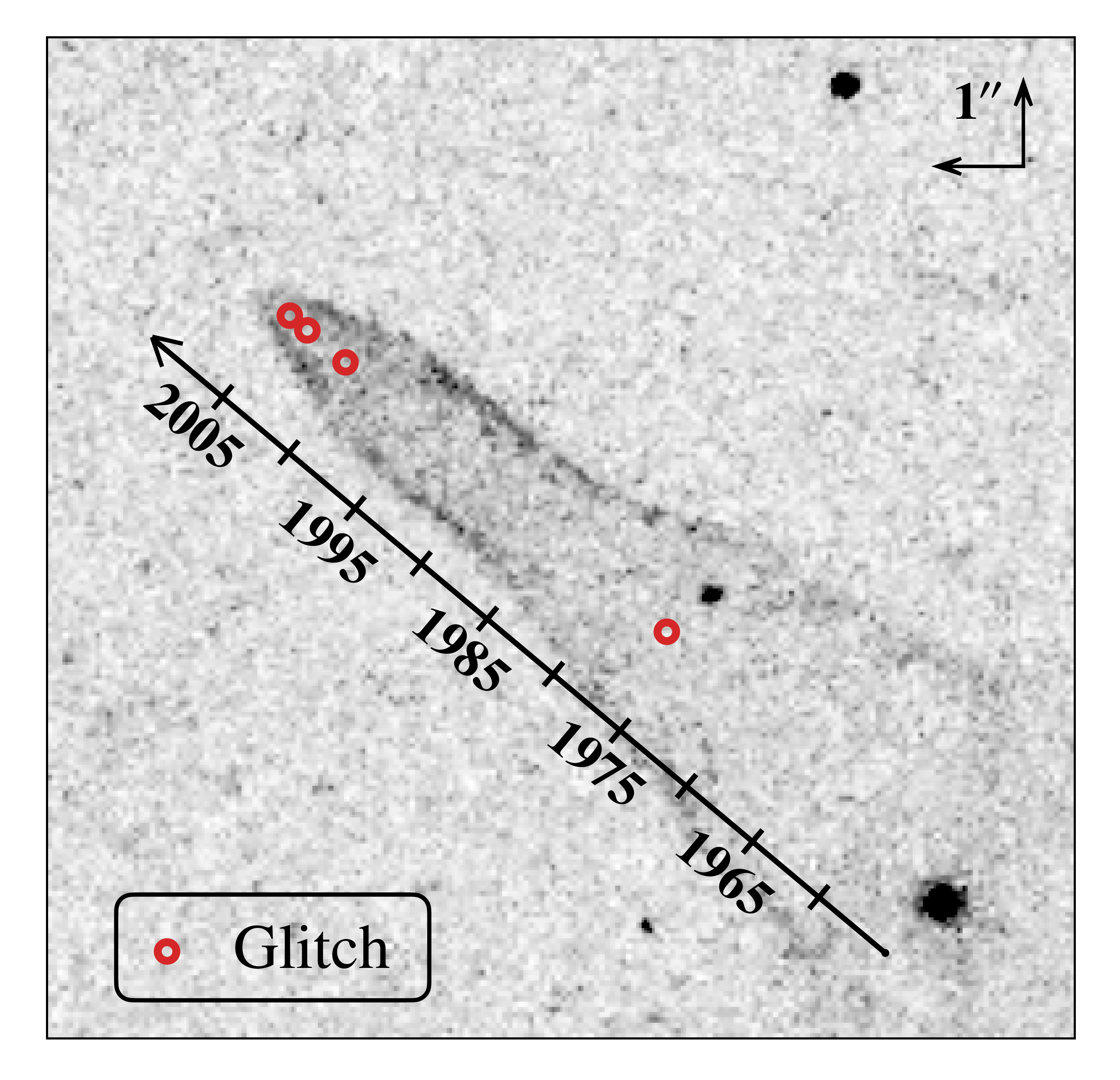}
    \caption{Locations of the pulsar during the four glitches reported between 1976 and 2005, overlaid as red open circles on the \textit{HST} image of the Guitar from 2006. The solid black line indicates the direction of the pulsar's proper motion, with tick marks indicating the location of the pulsar every five years along its historical trajectory. }
    \label{fig:Halphaglitch}
\end{figure}

\indent The fractional change $\Delta \dot{E}/\dot{E}$ and the corresponding change in stand-off radius $\Delta R_0/R_0$ during each glitch are shown in Table~\ref{tab:glitches}. The estimates of $\Delta R_0/R_0$ assume constant density during the glitch. Even for the largest glitch in 1976, the implied change in stand-off radius is only $0.1\%$, smaller than the spatial resolution of the H$\alpha$ images. The locations of the pulsar during each glitch are superimposed on the $HST$ image from 2006 in Figure~\ref{fig:Halphaglitch}. There is no discernible correlation between the locations of the glitches and the bow shock morphology. 

\indent \edit1{While we have only considered the change in spin-down luminosity in Table~\ref{tab:glitches}, it is unclear how much additional energy is processed through the pulsar wind during a glitch, and whether this additional energy could significantly alter the stand-off radius of a bow shock. Various magnetospheric effects have been observed during other pulsar glitches, including pulse shape changes in the Vela pulsar \citep{2018Natur.556..219P}, soft X-ray polarization changes in the Crab pulsar \citep{2020NatAs...4..511F}, and a pulse shape change in PSR B2021$+$51 \citep{2021ApJ...912...58L}. Such magnetospheric effects demonstrate that glitches may release additional energy that might impact the pulsar wind beyond the change in spin-down rate, but these effects are not a universal trait of glitches \citep{2018MNRAS.478.3832S, 2021MNRAS.504..406J} and have not been observed in any of the glitches from B2224$+$65.}

\indent \edit1{In the case of B2224$+$65, the only empirical handles on energy loss during its glitches are $\nu$ and $\dot{\nu}$. In addition to converting $\Delta \dot{E}/\dot{E}$ to $\Delta R_0/R_0$, as shown in Table~\ref{tab:glitches}, we can estimate the instantaneous change in spin energy during a glitch as $\Delta E \approx I \Omega \Delta \Omega \approx 2(\Delta \Omega/\Omega)E$, which is about $10^{-6}E\approx4\times10^{46}$ erg for the largest glitch in 1976. The glitch timescale is limited by the observation windows and is constrained to the time period between 13 September 1976 and 26 November 1976 \citep{1982ApJ...255L..63B}. Adopting 1.5 months as a conservative estimate of the glitch timescale, we have $\Delta \dot{E} \approx \Delta E/\Delta t \approx 10^{40}$ erg s$^{-1}$. If all of this energy were converted into the pulsar wind, then $\Delta R_0/R_0$ would be as large as 1000, and we would expect a substantial physical imprint of the glitch on the GN's structure. However, it is unclear if this spin energy is mediated through the pulsar wind or through other processes like electromagnetic radiation, and for the three most recent glitches, the implied glitch energies are far too small to cause any detectable change in the stand-off radius. Moreover, the energy release from glitches would most likely increase the stand-off radius, whereas the narrowing of the bow shock apex during the three most recent glitches indicates that the stand-off radius is decreasing. Glitches therefore appear to have a negligible impact on the observed morphology of the GN, suggesting that interstellar density fluctuations are likely the main driver of the stand-off radius variations inferred from images of the GN.}

\indent \edit1{However, we cannot rule out the general possibility that large glitches may impact the morphology of pulsar bow shocks, regardless of whether it appears to be a relevant effect for our analysis of the GN. For example, the quasi-periodically nulling pulsar B1931$+$24 exhibits a $50\%$ increase in its spin-down rate during 5-10 day-long active phases \citep{2006Sci...312..549K}. In this case, quasi-periodic variations in the stand-off radius would occur between the active and in-active phases. If this pulsar has an H$\alpha$ bow shock, then it would provide an ideal test bed for determining whether substantial changes in spin-down rate can be identified in large-scale bow shock images. We note however that even if B2224$+$65 experienced as extreme spin-down behaviors as B1931$+$24, the impact on the bow shock structure would likely be undetected because the timescale of the spin-down variations is much less than a year, the approximate timescale on which variations in the bow shock structure are resolved.}

\begin{deluxetable}{C C C C C}\label{tab:glitches}
\tablecaption{PSR B2224$+$65: Glitches}
\tabletypesize{\footnotesize}
\tablehead{\colhead{Year} & \colhead{$\Delta \nu/\nu$} & \colhead{$\Delta \dot{\nu}/\dot{\nu}$} & \colhead{$\Delta \dot{E}/\dot{E}$ } & \colhead{$\Delta R_0/R_0$} \\ \colhead{} & \colhead{($10^{-9}$)} & \colhead{($10^{-3}$)} & \colhead{($10^{-9}$)} & \colhead{} }
\startdata
1976^{\rm(1,2)} & 1.707(1)\times10^{3} & -3(5) & 1707 & 0.0013 \\
2000^{\rm(3)} & 0.14(3) & -2.9(2) & \nodata &  \nodata \\
2003^{\rm(3)} & 0.08(4) & -1.4(2) & \nodata & \nodata \\
2005^{\rm(3)} & 0.19(6) & \nodata & 0.19 & 1.4\times10^{-5} \\
2007^{\rm(4)} & 0.39(7) & -0.6(4) & 0.39 & 1.9\times10^{-5}\\
\enddata
\tablecomments{Left to right: Year and fractional changes in spin frequency $\Delta \nu/\nu$, spin frequency derivative $\Delta \dot{\nu}/\dot{\nu}$, spin-down luminosity $\Delta \dot{E}/\dot{E}$, and bow shock stand-off radius $\Delta R_0/R_0$ for each glitch. The changes in stand-off radius were calculated assuming constant density during the glitches.  Values of $\Delta \dot{E}/\dot{E}$ and $\Delta R_0/R_0$ are not shown for the 2000 and 2003 glitches because they do not have published pre- and post-glitch timing solutions. For the other glitches, $\Delta \dot{\nu}/\dot{\nu}$ was ignored due to its large uncertainty. References: (1) \cite{1982ApJ...255L..63B}, (2) \cite{1996MNRAS.282..677S}, (3)  \cite{2006AA...457..611J}, (4) \cite{2010MNRAS.404..289Y}.}
\end{deluxetable}

\subsection{Comparison to Other Pulsar Bow Shocks}\label{sec:othershocks}
At the time of this work, five other neutron stars with H$\alpha$ emitting bow shocks have been observed over multiple epochs spanning years to decades: J0437$-$4715 \citep{1993Natur.364..603B,1995ApJ...440L..81B, 2014ApJ...784..154B}, B0740$-$28 \citep{2002AA...389L...1J, 2014ApJ...784..154B}, J1741$-$2054 \citep{2010ApJ...724..908R, 2016ApJ...825..151M},  J2030$+$4415 \citep{2014ApJ...784..154B,2020ApJ...896L...7D}, and J2124$-$3358 \citep{2002ApJ...580L.137G, 2014ApJ...784..154B, 2017ApJ...851...61R}. Of these bow shocks, all but that of J0437$-$4715 show complex morphologies including closed bubbles, asymmetries, and undulating structures reminiscent of the Guitar (for a compilation of characteristic images, see \citealt{2014ApJ...784..154B}). The shock apex distances $\theta_a$ inferred by previous works are shown in Table~\ref{tab:psr}. Previous studies generally quote an empirically measured $\theta_a$ and assume an inclination angle $i = 90^\circ$, but some studies (\cite{2010ApJ...724..908R} for J1741$-$2054 and  \cite{2017ApJ...851...61R} for J2124$-$3358) perform more complicated fits for the stand-off radius and inclination angle simultaneously, and apex distances that are fit with different methods should not be considered as necessarily compatible even for the same bow shock. Various published images of each bow shock are also generally obtained from different telescopes and instruments, and hence vary in terms of resolution, seeing, and exposure time. Rather than re-analyze publicly available images of each bow shock in a self-consistent manner (which we relegate to future work), we adopt variations in the stand-off radii quoted from various studies as upper limits, and in Section~\ref{sec:spectrum} we calculate upper limits on the density wavenumber spectrum for each bow shock accordingly.

\section{Turbulence in the VLISM}\label{sec:VLISM}

\begin{figure*}
    \centering
    \includegraphics[width=\textwidth]{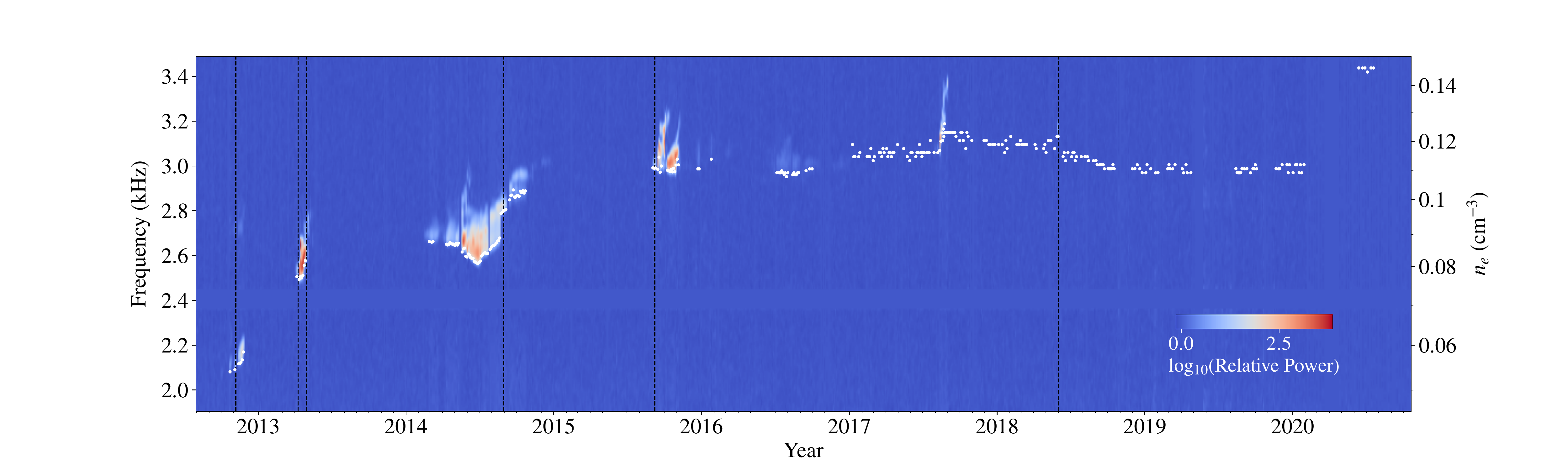}
    \caption{Dynamic spectrum showing all data from the Voyager 1 PWS wideband receiver since Voyager crossed the heliopause on August 25, 2012. The time resolution of the spectrum is 3 days and the frequency resolution is 0.011 kHz. Each column of pixels corresponds to a 1D spectrum that is the average of all 48-s epochs that fall within a given time bin, and which have been equilibrated to the same noise baseline. The power supply interference line at 2.4 kHz is masked, and data dropouts and periods of degraded telemetry are also masked. The spectrum is smoothed in frequency using a Gaussian kernel with $\sigma = 0.01$ kHz. Black dashed lines indicate epochs that were used to calculate density variations on 0.36-s timescales (see Figure~\ref{fig:highq_densities}), while the white points show average densities inferred from each epoch. The white points between 2017 and 2020 correspond to densities measured using techniques outlined in \cite{2021NatAs.tmp...84O}.}
    \label{fig:V1fullspec}
\end{figure*}

\subsection{Plasma Oscillations}
Voyager 1 (V1) measures the electron density of interstellar space by detecting plasma oscillations with the Plasma Wave System (PWS) \citep{2013Sci...341.1489G}. The PWS wideband receiver obtains voltage time series sampled at a rate of 28.8 kHz that are stored for later transmission to ground. These voltage time series are then converted to a frequency-time dynamic spectrum by Fourier methods, although automatic gain control on the wideband receiver prevents this spectrum from being calibrated to absolute electric field intensities. Nonetheless, the frequency of plasma oscillations detected in the PWS spectrum can be used to infer the plasma density. All of the data from the V1 PWS wideband receiver since V1's crossing of the heliopause in 2012 through late 2020 are shown in Figure~\ref{fig:V1fullspec}.

\indent Plasma oscillations are found in the PWS spectrum in two main ways. The first occurs when shocks from solar coronal mass ejections trigger plasma oscillation events (POEs) seen by V1 as brief, days to year-long bursts of power in the PWS spectrum, and these events often exhibit extended frequency structure and sharp monotonic increases in the plasma frequency that are associated with shocks passing over the spacecraft \citep{2015ApJ...809..121G,2021AJ....161...11G}. POEs have been detected by V1 approximately once per year since 2012, and their large intensities can allow the plasma frequency to be measured down to the smallest temporal resolution in the PWS spectrum, typically about 0.4 s. As V1 travels at a speed of about 17 km/s, the PWS spectrum can therefore, at least in theory, resolve density fluctuations on scales as small as about 7 kilometers. 

\indent The second class of plasma oscillations found by V1 are extremely weak, narrowband plasma oscillations that are detected in data from early 2017 through mid-2020 \citep{2021NatAs.tmp...84O}. While the physical origin of these \edit1{persistent} plasma waves is not entirely clear, they are detectable in the absence of POEs and do not appear to be associated with solar-origin shocks. The low signal-to-noise ratio (S/N) of the \edit1{persistent} plasma waves means that they are only detectable after averaging over at least one epoch of V1 PWS data, and hence they can be used to resolve density fluctuations on length scales as small as about 0.03 au. The combined density time series from both POEs and \edit1{persistent} plasma waves constrains the density fluctuation spectrum over wavenumbers from $10^{-12} \lesssim q \lesssim 10^{-3}$ m$^{-1}$.

\subsection{Plasma Frequency Measurements}
\subsubsection{High Wavenumber Regime}\label{sec:highq}
\begin{figure*}
    \centering
    \gridline{\fig{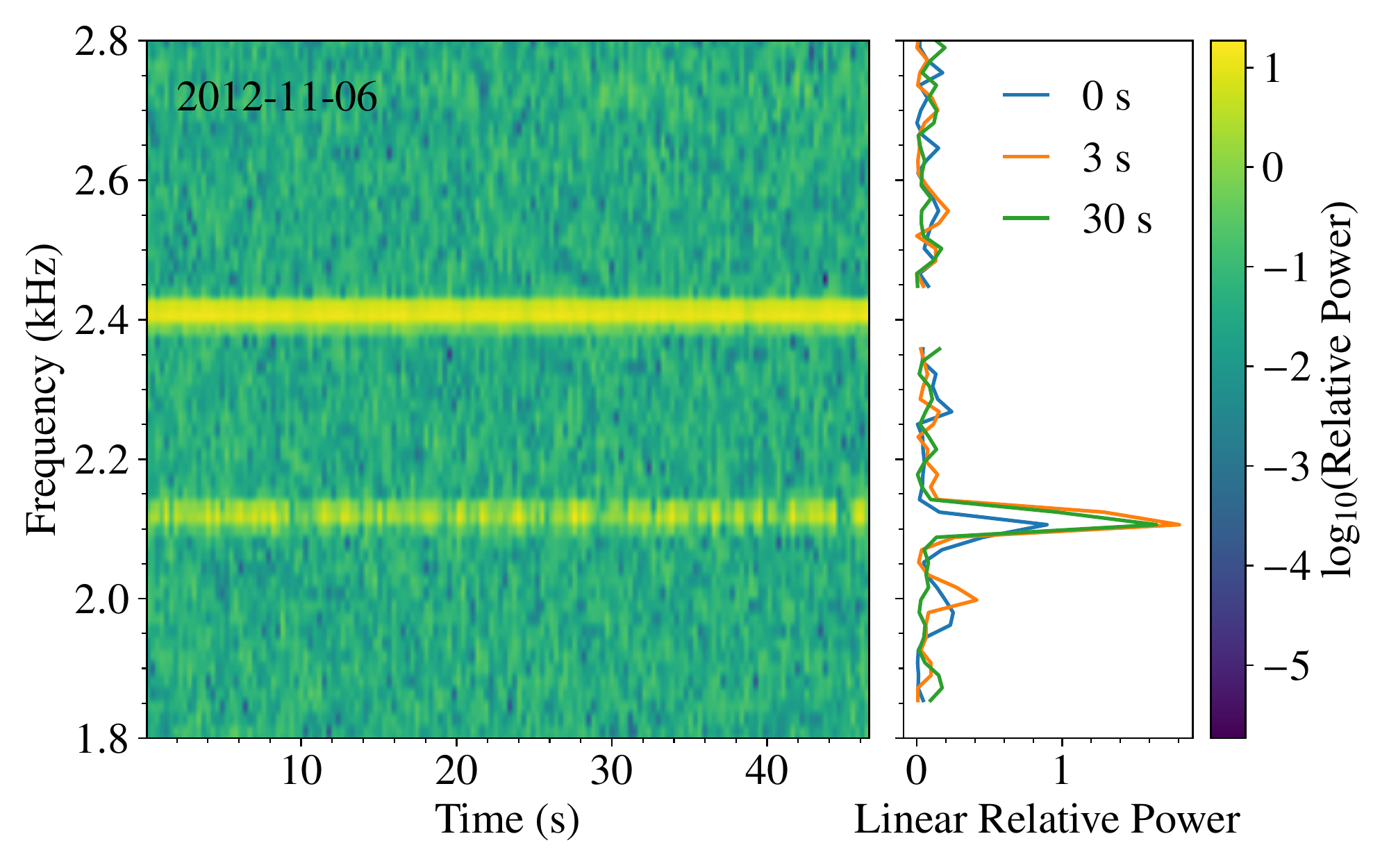}{0.48\textwidth}{(a)}
          \fig{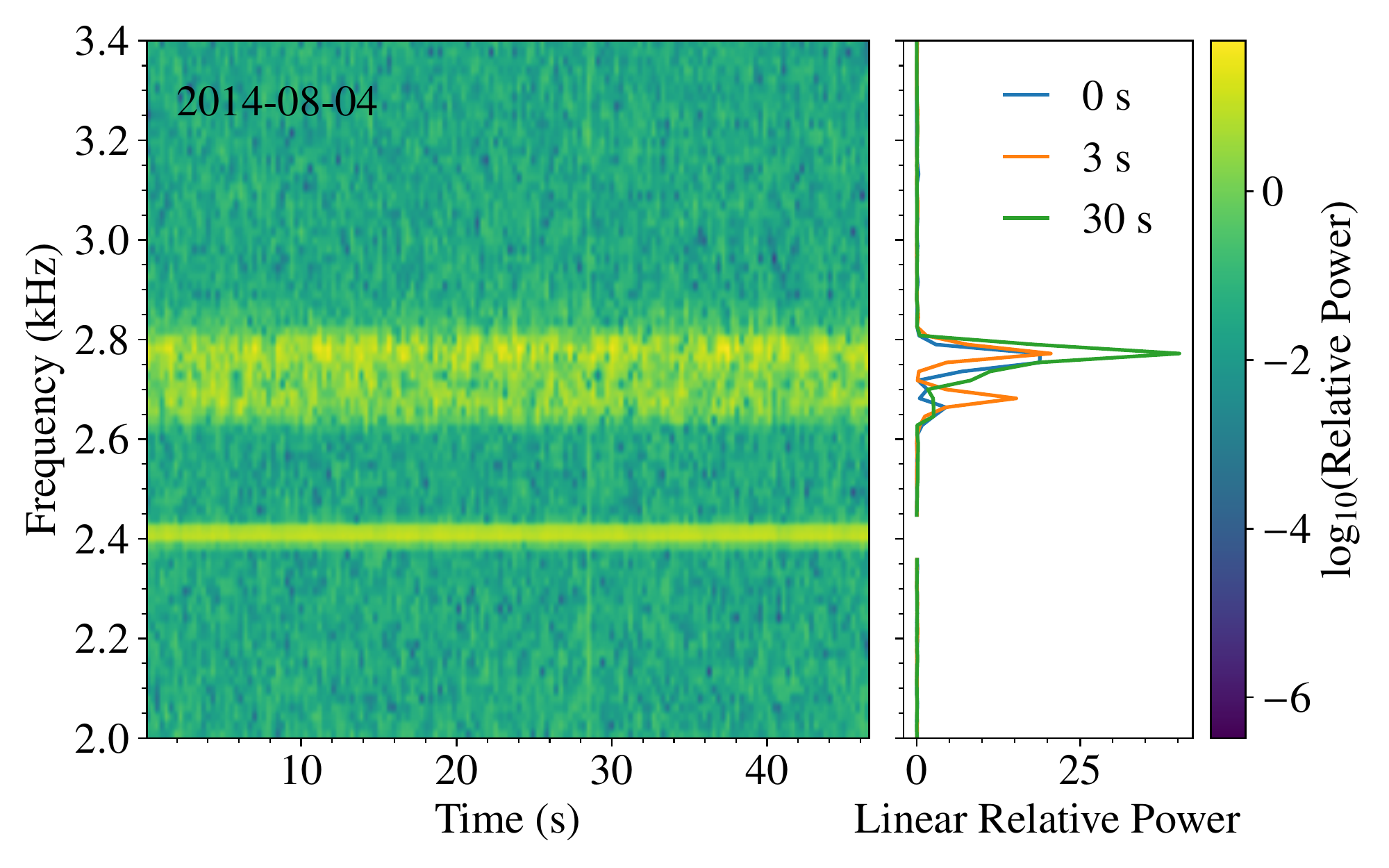}{0.48\textwidth}{(b)}}
    \caption{\textbf{(a)} A single epoch of V1 PWS data from November 6, 2012. The 2D dynamic spectrum has a temporal resolution of 0.36 s and a frequency resolution of 18 Hz, and it displays the plasma oscillation line at 2.1 kHz and the power supply interference line at 2.4 kHz. One dimensional slices through the spectrum at 0, 3, and 30 s are also shown. \textbf{(b)} A single epoch of V1 PWS data from August 4, 2014. The 2D dynamic spectrum displays plasma oscillations with frequency structure extending from 2.6 to 2.8 kHz that is associated with radio emission and frequency sidebands. One dimensional slices through the spectrum at 0, 3, and 30 seconds are also shown.}
    \label{fig:V1examples}
\end{figure*}

\begin{figure}
    \centering
    \includegraphics[width=0.5\textwidth]{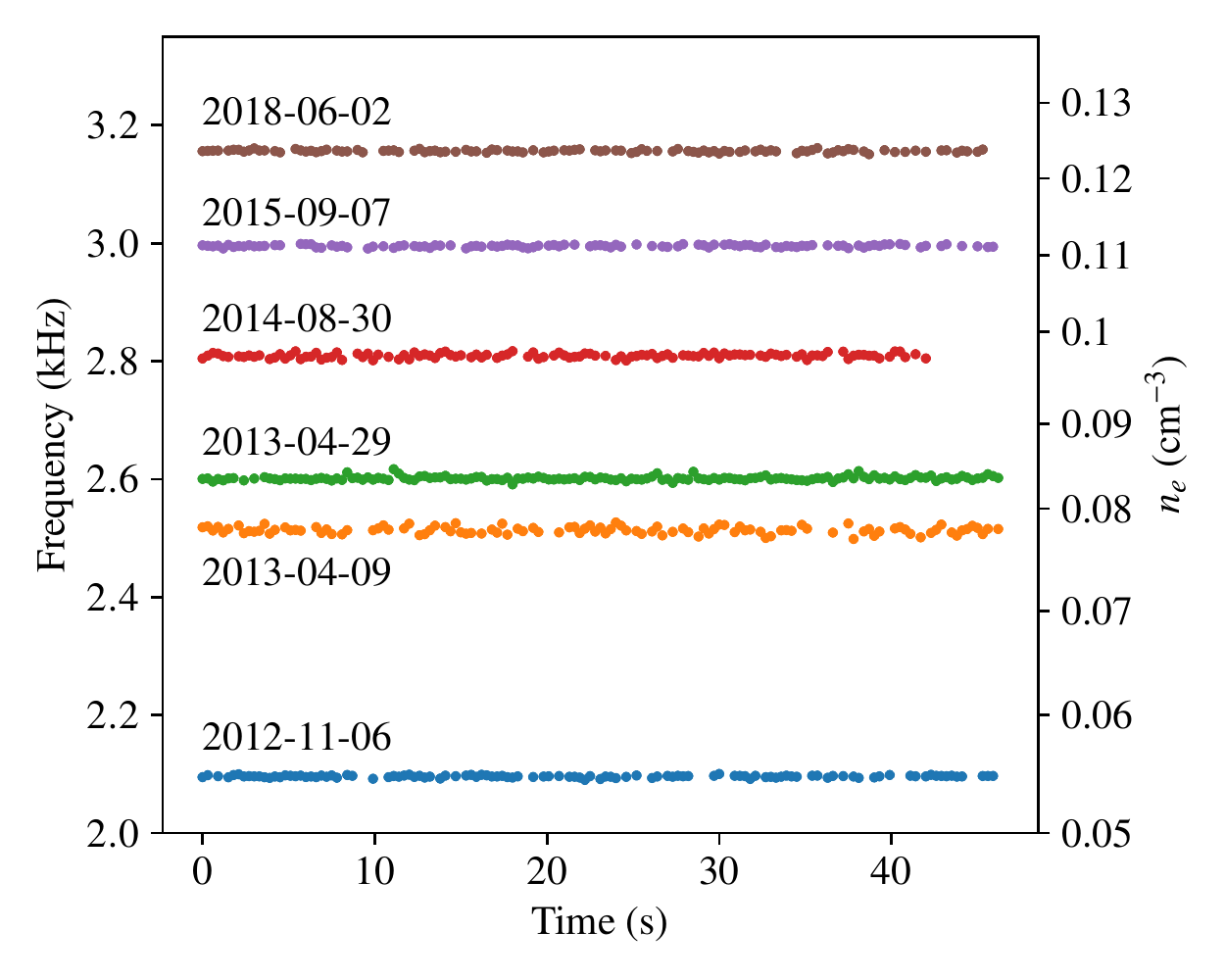}
    \caption{Plasma frequencies (left-hand axis) and densities (right-hand axis) measured for six epochs of V1 PWS data on short (0.36 s) timescales. Each epoch is about 48 s long, and gaps appear where the S/N was too low to accurately detect the plasma line. These epochs correspond to the black dashed lines in Figure~\ref{fig:V1fullspec} and were chosen based on the narrow bandwidth ($\approx 0.02$ to $0.04$ kHz) and high S/N of the plasma oscillations.}
    \label{fig:highq_densities}
\end{figure}

\indent In order to measure the plasma frequency over the smallest spatial scales, we extracted six single epochs of PWS data during POEs indicated by the black lines in Figure~\ref{fig:V1fullspec}. During these epochs, the plasma line had a narrow bandwidth ($\approx 0.05$ kHz) and a high S/N ($>10$), allowing for precise characterization of variations in the plasma frequency over the shortest timescales within a given epoch. An example of one of these epochs is shown in Figure~\ref{fig:V1examples}a. The plasma frequency was measured using a 1D matched filtering approach, where each column of the 2D spectrum was convolved with a Gaussian pulse in frequency space. The plasma frequency corresponds to the lower frequency cutoff of the plasma oscillations, which was taken to be the lower edge of the FWHM \citep[e.g.,][]{2019NatAs...3.1024G,2019NatAs...3..154L, 2020ApJ...904...66L}. Figure ~\ref{fig:V1examples}a also shows 1D slices through the spectrum at different times during the observation, demonstrating amplitude variations in the intensity of the plasma line. These amplitude variations do not impact the plasma frequency extraction due the narrow bandwidth and high S/N of the plasma line. The plasma frequencies and corresponding densities for all six epochs are shown in Figure~\ref{fig:highq_densities}. 

\indent While the POEs are generally detected within individual PWS epochs to high S/N, they typically exhibit a range of complex variability that makes it difficult to differentiate between plasma frequency variations related to changes in the underlying density and variations associated with wave interactions or the instrumental quantization of the signal. Almost every POE displays a combination of plasma oscillation sidebands, which are typically attributed to Langmuir parametric decay and the excitation of higher wave modes, and trapped radio emissions that augment the plasma line \citep[e.g.,][]{2013Sci...341.1489G}. In many cases, the plasma oscillation sidebands vary in intensity within a single epoch, making it difficult to accurately track the lowest frequency sideband that would be attributed to the plasma frequency. Epochs where the plasma line is dominated by broadband radio emissions also display a large degree of variability over the shortest timescales, and it is unclear whether apparent frequency variations during these epochs are an accurate measure of the underlying density. A typical example of an epoch containing broadband radio emission is shown in Figure~\ref{fig:V1examples}b. Instrumental quantization further complicates interpretation of the apparent variability in these epochs, as the finite frequency resolution of the spectrum can effectively smear out or even enhance apparent frequency variations when the plasma line contains sidebands or broadband radio emission. By contrast, the six epochs chosen to constrain density variations on the smallest spatial scales are characterized by narrow plasma lines that do not contain any evidence of frequency sidebands or broadband emission. The narrow bandwidth of the plasma line during these epochs mitigates the signal quantization, but we nonetheless interpret the plasma frequency variations in these epochs as upper limits on the underlying, turbulent density variations.

\subsubsection{Low Wavenumber Regime}

\indent Despite the complex variability displayed in most epochs containing POEs, the shape of the plasma line is generally stable when averaged over a full 48 s epoch. Therefore, to measure density variations in the low wavenumber regime and over the largest spatial scales, we extract the plasma frequency by averaging every epoch in time to obtain a 1D spectrum that is then analyzed through the same matched filtering technique applied to the individual epochs described in Section~\ref{sec:highq}. The resulting densities are overlaid on the full PWS spectrum in Figure~\ref{fig:V1fullspec}. We also extract the persistent, narrowband plasma waves apparent in the PWS spectrum beginning in early 2017 using the same techniques outlined in \cite{2021NatAs.tmp...84O}. Our analysis includes newer data extending through October 2020, when the frequency of the \edit1{persistent} plasma line increases by a factor of about 1.1 due to the passage of a magnetic pressure front over the spacecraft \citep{2021ApJ...911...61B}. The densities extracted from the \edit1{persistent} plasma waves are also shown in Figure~\ref{fig:V1fullspec}. We ignore several epochs bordering shock discontinuities in the 2014 and 2017 POEs because these discontinuities do not reflect a turbulent process. 

\section{Results: The Composite Wavenumber Spectrum}\label{sec:spectrum}

\begin{figure}
    \centering
    \includegraphics[width=0.48\textwidth]{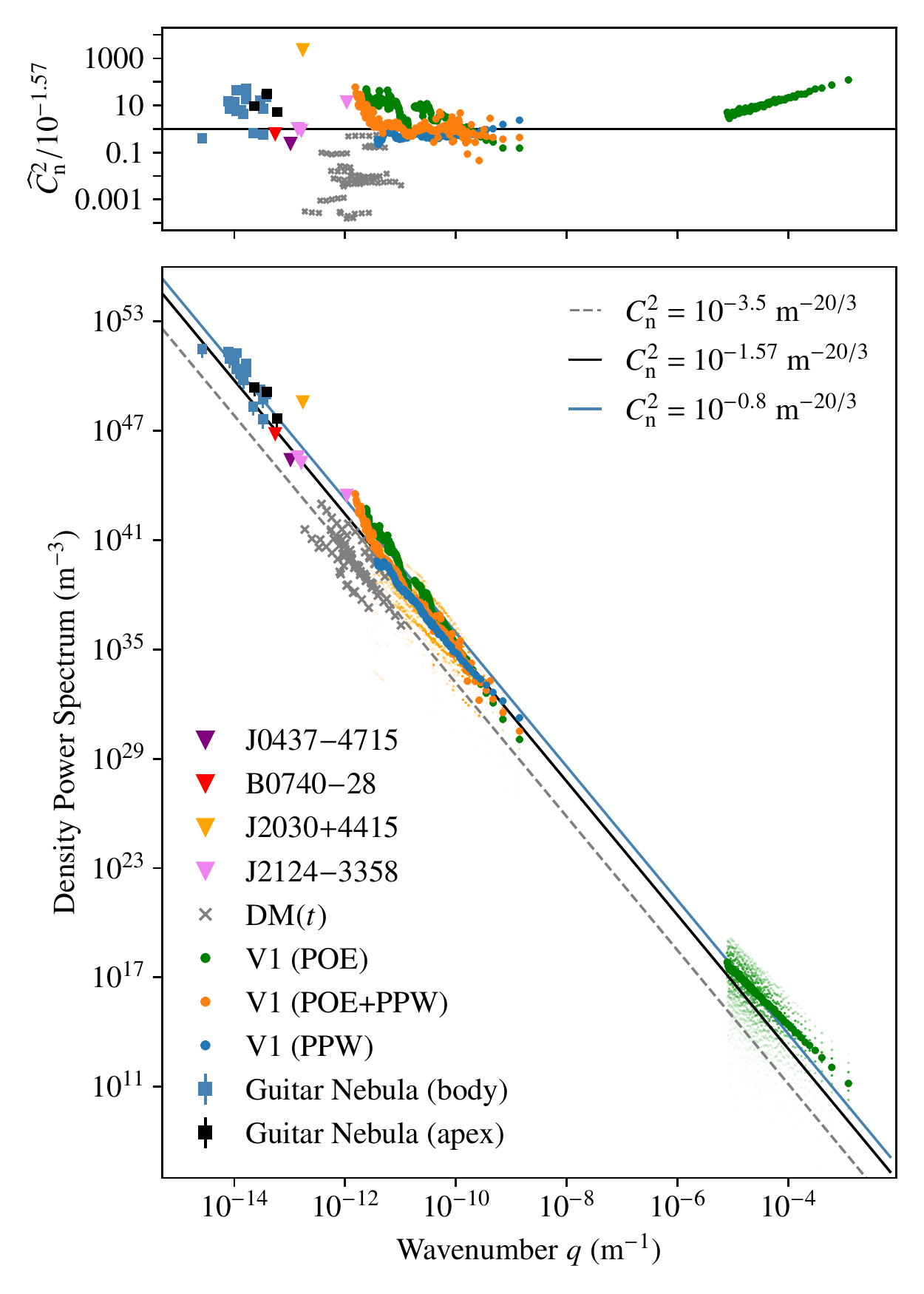}
    \caption{\textbf{Bottom:} The 3D wavenumber spectrum of density fluctuations inferred from both in situ and integrated density measurements. The spectrum inferred from observations of the Guitar Nebula is shown in black for densities that were obtained using only the shock apex, where the \cite{1996ApJ...459L..31W} provides a more precise constraint on the stand-off radius, and in light blue for densities that were obtained by extrapolating the \cite{1996ApJ...459L..31W} to various parts of the downstream shock. Squares with error bars for the Guitar indicate the range of \replaced{$C_{\rm n}^2$}{$\widehat{C}_{\rm n}^2$} for $i = 90^\circ\pm30^\circ$. Triangles indicate spectral constraints based on four other pulsar bow shocks: J0437$-$4715, B0740$-$28, J2030$+$4415, and J2124$-$3358; these spectral amplitudes are presented as upper limits (see Section~\ref{sec:othershocks}). The spectrum of densities obtained with Voyager 1 (V1) are colored according to the type of plasma wave phenomena used: green for plasma oscillation events (POEs), blue for \edit1{persistent} plasma waves \edit1{(PPW)} \edit1{spanning 2017 to 2020}, and orange for a combination of POEs and the \edit1{persistent} plasma wave data. The spectral amplitudes shown for V1 are binned in wavenumber space and then averaged, but the full distribution of amplitudes is shown by the smaller, translucent green and orange points. Constraints from pulsar DM variations over time (DM$(t)$) are shown as grey crosses. The solid lines indicate the best-fit spectral amplitudes for the Guitar Nebula (light blue) and V1 data (black). The dashed line indicates a constant amplitude $C_{\rm n}^2 = 10^{-3.5}$ m$^{-20/3}$. \textbf{Top:} All estimated spectral amplitudes divided by the best-fit value for the VLISM probed by V1, $C_{\rm n}^2 \approx 10^{-1.57}$ m$^{-20/3}$. }
    \label{fig:wavenumber_spec}
\end{figure}

\indent The electron density measurements obtained from the GN and V1 PWS are used to calculate the wavenumber spectrum of density fluctuations using the analytic formalism described in Section~\ref{sec:theory} and fixing the spectral index to $\beta = 11/3$. The resulting spectrum is shown in Figure~\ref{fig:wavenumber_spec}. The spectral amplitudes obtained by fitting the Wilkin model to the tip of the GN and to various sections of the downstream shock are shown separately in Figure~\ref{fig:wavenumber_spec} and are consistent with a \added{mean} spectral amplitude $C_{\rm n}^2 = 10^{-0.8\pm0.2}$ m$^{-20/3}$. Figure~\ref{fig:wavenumber_spec} also shows upper limits on \added{point estimates of} \replaced{$C_{\rm n}^2$}{$\widehat{C}_{\rm n}^2$} for four other pulsar bow shocks described in Section~\ref{sec:othershocks}. 

\indent Owing to the temporal resolution and sampling of the V1 data on both short and long timescales, the spectrum of V1 densities falls into two wavenumber regimes: $q\gtrsim 10^{-5}$ m$^{-1}$ and $10^{-12} \lesssim q \lesssim 10^{-8}$ m$^{-1}$. In both cases, the density structure function was binned in wavenumber space and then averaged to obtain a single constraint \replaced{$C_{\rm n}^2$}{$\widehat{C}_{\rm n}^2$} per wavenumber bin. In the low wavenumber regime, density measurements are obtained from both POEs and persistent, narrowband plasma waves. The wavenumber spectrum obtained solely from the \edit1{persistent} plasma waves yields $C_{\rm n}^2 \approx 10^{-1.6}$ m$^{-20/3}$ \citep{2021NatAs.tmp...84O}, and when we include POE densities, the best-fit spectral amplitude obtained through nonlinear least squares fitting is $C_{\rm n}^2 = 10^{-1.57 \pm 0.02}$ m$^{-20/3}$; this fit ignores $q< 1.5\times10^{-12}$ m$^{-1}$ owing to a spectral excess discussed later. The spectral amplitude errors for both the Guitar and V1 are only based on the fit for $C_{\rm n}^2$ from the data, and do not account for statistical variations associated with sampling a single realization of a Kolmogorov process. We therefore interpret the spectral amplitude errors for both the Guitar and V1 as lower limits on the true uncertainties.

\indent Our constraint on $C_{\rm n}^2$ for the VLISM probed by V1 is consistent with the \cite{2020ApJ...904...66L} study that examined V1 data extending through June 2019, and which found $C_{\rm n}^2 = 10^{-1.47\pm0.04}$ m$^{-20/3}$. Similar to \cite{2019NatAs...3..154L, 2020ApJ...904...66L}, we also find a power excess with a shallower spectral slope in the high wavenumber regime of the V1 spectrum. \cite{2019NatAs...3..154L, 2020ApJ...904...66L} suggest that this spectral excess at high wavenumbers may be associated with local kinetic wave activity that is triggered by the shocks responsible for POEs. It is possible that kinetic Alfv\'en waves are responsible for density fluctuations in the high wavenumber regime, and hence the observed spectral excess could be indicative of the underlying physical processes that are transmitting turbulence at these scales. However, constraints on the high wavenumber regime with V1 are ultimately limited by the finite resolution of the data in both time and frequency, and it remains possible that quantization of the PWS data on the shortest timescales may bias the observed density variations at the highest wavenumbers. We therefore interpret the amplitude of the V1 density spectrum as an upper limit in this high wavenumber regime.

\indent A power excess is also found at the lowest wavenumbers ($q<1.5\times10^{-12}$ m$^{-1}$) of the V1 spectrum, and is likely related to density variations over the largest spatial scales in the V1 data. At these scales, the observed density variations are influenced by a combination of turbulent and deterministic processes, like the discrete shocks that trigger POEs and cause density jumps in the PWS spectrum. It can be difficult to disentangle solar-origin shocks from the underlying structure of the VLISM, which is largely determined by interactions between the interstellar plasma and magnetic fields with those of the solar wind. For example, between 2013 and 2015 multiple density jumps are observed in the PWS spectrum, two of which are directly associated with shock waves observed in the V1 magnetic field data \citep{2013ApJ...778L...3B, 2015ApJ...809..121G}. However, the rise in density between 2013 and 2015 also appears to be a persistent, structural feature of the VLISM, as the density remains roughly constant from 2015 through early 2020, when another magnetic pressure wave and density jump are observed \citep{2021ApJ...911...61B}. While we ignore two well-defined shock discontinuities when calculating the V1 wavenumber spectrum, the spectral excess at low wavenumbers is likely still biased by discrete, structural variations in the plasma between the heliopause and Solar System bow shock/wave.

\indent The spectra of density fluctuations observed near the GN, the other four pulsar bow shocks examined, and in the VLISM are all consistent with a turbulence spectrum that is enhanced when compared to other pulsars' LOS through the local ISM. In Figure~\ref{fig:wavenumber_spec} we show constraints on the wavenumber spectrum from DM variations observed in the Nanohertz Observatory for Gravitational Waves (NANOGrav) 12.5 year data set \citep{2021ApJS..252....4A}.  The DM time series of 18 pulsars were chosen based on consistency of the DM structure functions with a Kolmogorov process, and the spectral amplitudes were then calculated from the DM structure functions according to Equation~\ref{eq:DMCn2}. Small temporal lags in the DM structure functions were ignored due to biasing from white noise, and large temporal lags were similarly ignored due to biasing from the finite length of the data set. The resulting structure functions constrain $C_{\rm n}^2$ over wavenumbers $10^{-13} \lesssim q \lesssim 10^{-11}$ m$^{-1}$. The DM variations are broadly consistent with $C_{\rm n}^2 \approx 10^{-3.5}$ m$^{-20/3}$, the typical value that has been found in previous studies of the turbulence spectrum in the warm ionized medium (WIM; e.g., \citealt{1995ApJ...443..209A, 2010ApJ...710..853C}). The DM$(t)$ wavenumber spectrum also exhibits an orders of magnitude spread in $C_{\rm n}^2$ that reflects large variations between different LOS through the local ISM.

\section{Discussion}\label{sec:conc}

\indent We present electron density fluctuation measurements obtained from H$\alpha$ images of the GN and from V1 PWS and constrain the density wavenumber spectrum over spatial scales of kilometers to 1000s of au. The characteristic electron densities in these regions are $n_e\sim0.3$ cm$^{-3}$ for the GN and $n_e\sim0.1$ cm$^{-3}$ for the VLISM. Comparison to previous observations of four other pulsar bow shocks show that all of the bow shocks examined in this study exhibit a spectral amplitude $C_{\rm n}^2$ that is orders of magnitude larger than values that are considered typical in the WIM. For the GN, we find $C_{\rm n}^2 = 10^{-0.8\pm0.2}$ m$^{-20/3}$, and for the VLISM probed by V1 we find $C_{\rm n}^2 = 10^{-1.57\pm0.02}$ m$^{-20/3}$. 

\indent It has already been suggested that the large value of $C_{\rm n}^2$ in the VLISM is the result of a turbulence spectrum that is enhanced by the superposition of interstellar and solar wind turbulence \citep{2019ApJ...887..116Z}. If that is the case, it is unclear how far Voyager would need to travel to sample ``pristine" or ``quiescent" interstellar turbulence, although it is likely that Voyager would need to cross the heliospheric bow shock/wave. Nonetheless, beyond the heliospheric bow wave lies a collection of interstellar clouds (one of which encases the Solar System; \citealt{2019ApJ...886...41L}), and cloud-cloud interactions may further modify turbulent density variations in this region \citep{2004ApJ...613.1004R}. It is unclear whether interactions between a bow shock and the ISM lead to similarly large values of $C_{\rm n}^2$ for the GN and the other bow shocks considered in this work. We find no empirical evidence that discrete events from B2224$+$65, such as glitches, have any observable impact on the GN's structure and inferred density fluctuations. Moreover, the large range of $C_{\rm n}^2$ that we estimate from DM variations along a number of pulsar LOS suggests that apparent enhancements to $C_{\rm n}^2$ for the VLISM and the pulsar bow shocks considered may simply reflect larger-scale variability between different regions of the ionized ISM.

\indent It is also possible that H$\alpha$ detections of pulsar bow shocks are systematically biased towards regions of higher density, leading to a higher $C_{\rm n}^2$ than expected from pulsar DMs and scattering measurements. Pulsar bow shocks will only produce visible H$\alpha$ emission when the neutral fraction in the surrounding gas is large enough, although exactly how large it needs to be depends on other factors like the pulsar velocity and distance \citep{2002ApJ...575..407C, 2014ApJ...784..154B}. Generally speaking, the number densities of cold and warm neutral gas in the ISM are much larger ($n_{\rm H}\sim 30$ and $\sim 0.6$ cm$^{-3}$, respectively) than densities characteristic of the WIM ($\sim 0.01$ cm$^{-3}$) \citep{2011piim.book.....D}. The pulsars producing these bow shocks may also preionize the atomic gas, leading to a higher electron density \citep{2016ApJ...821...66L}. 

\indent In this case, linear DM variations may also be detected from the pulsars due to their motions away from or towards the observer, combined with discrete changes in DM caused by pre-ionization of gas ahead of the shock and the pulsars moving through gas of varying density. If this is the case, shock-induced DM variations will largely be detected from pulsars residing in atomic gas, which comprises about $60\%$ of the ISM \citep{2011piim.book.....D}. Recent analysis of the NANOGrav 9-year data set by \cite{2017ApJ...841..125J} found linear DM trends in 14 out of the 37 pulsars analyzed ($38\%$), but combinations of linear trends and other fluctuations were also found in an additional 14 of the 37 pulsars. It is possible that some of these linear DM variations are related to pulsar bow shocks residing in atomic gas. In our analysis, we focus on stochastic DM variations that are easily attributable to turbulent density fluctuations. DM variations are only sensitive to free electrons and are integrated over 10s of pc to kpc-scale distances, and in the absence of large-scale, discrete structures along the LOS, stochastic DM variations will generally trace the more diffuse WIM. We could therefore interpret the broad range of $C_{\rm n}^2$ estimated from the stochastic DM variations analyzed in this study as a reflection of a WIM that varies in structure and is permeated by HII regions and bubbles, whereas estimates of $C_{\rm n}^2$ based on pulsar bow shocks may generally trace denser, more neutral media. If this is the case, then continued comparison of density fluctuations from direct observations of bow shocks and remote observations of pulsar DM and scattering may allow us to distinguish the properties of turbulence for a range of gas conditions in the ISM. 

\indent Previous studies on the density and magnetic field power spectra of interstellar turbulence in the ionized components of the ISM have generally focused on consistency with a Kolmogorov spectral index over many decades in wavenumber. In this study, we aim to call attention to departures from a uniform turbulence spectrum. Not only do we find large variability in the spectral amplitude between different pulsars' LOS through the ISM, but we also find significant enhancements in the spectral amplitude near the Solar System bow shock/wave and for the pulsar bow shocks considered. It is unclear whether these enhanced spectral amplitudes are characteristic of stellar bow shocks in general, and hence represent some local feature of turbulence in these environments. Given that the ISM is permeated by stars emitting winds and flares, high-velocity stars driving shocks, and supernovae, the mechanisms by which turbulence is mediated through these various phenomena is of high interest. Previous studies have already demonstrated that additional properties of the turbulence spectrum, such as the outer and inner scales, sonic regime, and spectral amplitude and slope, may vary between different regions of the local ISM and across the Galaxy \citep[e.g.,][]{1985ApJ...288..221C, 2015ApJ...804...23K}. In the VLISM, the outer scale constrained by magnetic field and density fluctuations is about 0.01 pc \citep{2018ApJ...854...20B, 2020ApJ...904...66L}, whereas in the Galactic thick disk the outer scale may be as large as 100 pc, and will generally depend on the mechanisms that drive turbulence in a particular region, such as stellar winds and supernovae. Similarly, the inner scale may vary depending on the local magnetic field strength and resulting proton gyroradius. In future work, we will assess variations in the density wavenumber spectrum in the context of pulsar scattering measurements and their spatial distribution, and evaluate these variations with respect to previous studies examining systematic differences between turbulence in the inner and outer Galaxy. 

\acknowledgements{S.K.O., J.M.C., and S.C. acknowledge support from the National Aeronautics and Space Administration (NASA 80NSSC20K0784). The authors also acknowledge support from the National Science Foundation (NSF AAG-1815242) and are members of the NANOGrav Physics Frontiers Center, which is supported by the NSF award PHY-1430284. This work is based in part on observations made with the NASA/ESA Hubble Space Telescope, obtained at the Space Telescope Science Institute, which is operated by the Association of Universities for Research in Astronomy, Inc., under NASA contract NAS5-26555. These observations are associated with programs 5387, 9129, and 10763. The NANOGrav 12.5 year data set contains observations from the Arecibo, Green Bank, and National Radio Astronomy Observatories. The Arecibo Observatory is a facility of the NSF operated under cooperative agreement (\#AST-1744119) by the University of Central Florida (UCF) in alliance with Universidad Ana G. Méndez (UAGM) and Yang Enterprises (YEI), Inc. The Green Bank Observatory and National Radio Astronomy Observatory are facilities of the NSF operated under cooperative agreement by Associated Universities, Inc.} T.D. acknowledges NSF AAG award number 2009468.

\bibliography{bib}

\begin{thebibliography}{}
\expandafter\ifx\csname natexlab\endcsname\relax\def\natexlab#1{#1}\fi
\providecommand{\url}[1]{\href{#1}{#1}}
\providecommand{\dodoi}[1]{doi:~\href{http://doi.org/#1}{\nolinkurl{#1}}}
\providecommand{\doeprint}[1]{\href{http://ascl.net/#1}{\nolinkurl{http://ascl.net/#1}}}
\providecommand{\doarXiv}[1]{\href{https://arxiv.org/abs/#1}{\nolinkurl{https://arxiv.org/abs/#1}}}

\bibitem[{{Alam} {et~al.}(2021){Alam}, {Arzoumanian}, {Baker}, {Blumer},
  {Bohler}, {Brazier}, {Brook}, {Burke-Spolaor}, {Caballero}, {Camuccio},
  {Chamberlain}, {Chatterjee}, {Cordes}, {Cornish}, {Crawford}, {Cromartie},
  {Decesar}, {Demorest}, {Dolch}, {Ellis}, {Ferdman}, {Ferrara}, {Fiore},
  {Fonseca}, {Garcia}, {Garver-Daniels}, {Gentile}, {Good}, {Gusdorff},
  {Halmrast}, {Hazboun}, {Islo}, {Jennings}, {Jessup}, {Jones}, {Kaiser},
  {Kaplan}, {Kelley}, {Key}, {Lam}, {Lazio}, {Lorimer}, {Luo}, {Lynch},
  {Madison}, {Maraccini}, {McLaughlin}, {Mingarelli}, {Ng}, {Nguyen}, {Nice},
  {Pennucci}, {Pol}, {Ramette}, {Ransom}, {Ray}, {Shapiro-Albert}, {Siemens},
  {Simon}, {Spiewak}, {Stairs}, {Stinebring}, {Stovall}, {Swiggum}, {Taylor},
  {Tripepi}, {Vallisneri}, {Vigeland}, {Witt}, {Zhu}, \& {Nanograv
  Collaboration}}]{2021ApJS..252....4A}
{Alam}, M.~F., {Arzoumanian}, Z., {Baker}, P.~T., {et~al.} 2021, ApJS, 252, 4,
  \dodoi{10.3847/1538-4365/abc6a0}

\bibitem[{{Aldcroft} {et~al.}(1992){Aldcroft}, {Romani}, \&
  {Cordes}}]{1992ApJ...400..638A}
{Aldcroft}, T.~L., {Romani}, R.~W., \& {Cordes}, J.~M. 1992, ApJ, 400, 638,
  \dodoi{10.1086/172025}

\bibitem[{{Armstrong} {et~al.}(1995){Armstrong}, {Rickett}, \&
  {Spangler}}]{1995ApJ...443..209A}
{Armstrong}, J.~W., {Rickett}, B.~J., \& {Spangler}, S.~R. 1995, ApJ, 443, 209,
  \dodoi{10.1086/175515}

\bibitem[{{Backus} {et~al.}(1982){Backus}, {Taylor}, \&
  {Damashek}}]{1982ApJ...255L..63B}
{Backus}, P.~R., {Taylor}, J.~H., \& {Damashek}, M. 1982, ApJL, 255, L63,
  \dodoi{10.1086/183770}

\bibitem[{{Barkov} {et~al.}(2020){Barkov}, {Lyutikov}, \&
  {Khangulyan}}]{2020MNRAS.497.2605B}
{Barkov}, M.~V., {Lyutikov}, M., \& {Khangulyan}, D. 2020, MNRAS, 497, 2605,
  \dodoi{10.1093/mnras/staa1601}

\bibitem[{{Bell} {et~al.}(1993){Bell}, {Bailes}, \&
  {Bessell}}]{1993Natur.364..603B}
{Bell}, J.~F., {Bailes}, M., \& {Bessell}, M.~S. 1993, Nature, 364, 603,
  \dodoi{10.1038/364603a0}

\bibitem[{{Bell} {et~al.}(1995){Bell}, {Bailes}, {Manchester}, {Weisberg}, \&
  {Lyne}}]{1995ApJ...440L..81B}
{Bell}, J.~F., {Bailes}, M., {Manchester}, R.~N., {Weisberg}, J.~M., \& {Lyne},
  A.~G. 1995, ApJL, 440, L81, \dodoi{10.1086/187766}

\bibitem[{{Brownsberger} \& {Romani}(2014)}]{2014ApJ...784..154B}
{Brownsberger}, S., \& {Romani}, R.~W. 2014, ApJ, 784, 154,
  \dodoi{10.1088/0004-637X/784/2/154}

\bibitem[{{Bucciantini}(2002)}]{2002AA...393..629B}
{Bucciantini}, N. 2002, A\&A, 393, 629, \dodoi{10.1051/0004-6361:20020968}

\bibitem[{{Burlaga} {et~al.}(2018){Burlaga}, {Florinski}, \&
  {Ness}}]{2018ApJ...854...20B}
{Burlaga}, L.~F., {Florinski}, V., \& {Ness}, N.~F. 2018, ApJ, 854, 20,
  \dodoi{10.3847/1538-4357/aaa45a}

\bibitem[{{Burlaga} {et~al.}(2021){Burlaga}, {Kurth}, {Gurnett},
  {Berdichevsky}, {Jian}, {Ness}, {Park}, \& {Szabo}}]{2021ApJ...911...61B}
{Burlaga}, L.~F., {Kurth}, W.~S., {Gurnett}, D.~A., {et~al.} 2021, ApJ, 911,
  61, \dodoi{10.3847/1538-4357/abeb6a}

\bibitem[{{Burlaga} {et~al.}(2013){Burlaga}, {Ness}, {Gurnett}, \&
  {Kurth}}]{2013ApJ...778L...3B}
{Burlaga}, L.~F., {Ness}, N.~F., {Gurnett}, D.~A., \& {Kurth}, W.~S. 2013,
  ApJL, 778, L3, \dodoi{10.1088/2041-8205/778/1/L3}

\bibitem[{{Chatterjee} \& {Cordes}(2002)}]{2002ApJ...575..407C}
{Chatterjee}, S., \& {Cordes}, J.~M. 2002, ApJ, 575, 407,
  \dodoi{10.1086/341139}

\bibitem[{{Chatterjee} \& {Cordes}(2004)}]{2004ApJ...600L..51C}
---. 2004, ApJL, 600, L51, \dodoi{10.1086/381498}

\bibitem[{{Chepurnov} \& {Lazarian}(2010)}]{2010ApJ...710..853C}
{Chepurnov}, A., \& {Lazarian}, A. 2010, ApJ, 710, 853,
  \dodoi{10.1088/0004-637X/710/1/853}

\bibitem[{{Cordes} \& {Lazio}(2002)}]{2002astro.ph..7156C}
{Cordes}, J.~M., \& {Lazio}, T.~J.~W. 2002, arXiv e-prints, astro.
\newblock \doarXiv{astro-ph/0207156}

\bibitem[{{Cordes} {et~al.}(1993){Cordes}, {Romani}, \&
  {Lundgren}}]{1993Natur.362..133C}
{Cordes}, J.~M., {Romani}, R.~W., \& {Lundgren}, S.~C. 1993, Nature, 362, 133,
  \dodoi{10.1038/362133a0}

\bibitem[{{Cordes} {et~al.}(1985){Cordes}, {Weisberg}, \&
  {Boriakoff}}]{1985ApJ...288..221C}
{Cordes}, J.~M., {Weisberg}, J.~M., \& {Boriakoff}, V. 1985, ApJ, 288, 221,
  \dodoi{10.1086/162784}

\bibitem[{{de Vries} \& {Romani}(2020)}]{2020ApJ...896L...7D}
{de Vries}, M., \& {Romani}, R.~W. 2020, ApJL, 896, L7,
  \dodoi{10.3847/2041-8213/ab9640}

\bibitem[{{Decin} {et~al.}(2012){Decin}, {Cox}, {Royer}, {Van Marle},
  {Vandenbussche}, {Ladjal}, {Kerschbaum}, {Ottensamer}, {Barlow}, {Blommaert},
  {Gomez}, {Groenewegen}, {Lim}, {Swinyard}, {Waelkens}, \&
  {Tielens}}]{2012A&A...548A.113D}
{Decin}, L., {Cox}, N.~L.~J., {Royer}, P., {et~al.} 2012, A\&A, 548, A113,
  \dodoi{10.1051/0004-6361/201219792}

\bibitem[{{Deller} {et~al.}(2008){Deller}, {Verbiest}, {Tingay}, \&
  {Bailes}}]{2008ApJ...685L..67D}
{Deller}, A.~T., {Verbiest}, J.~P.~W., {Tingay}, S.~J., \& {Bailes}, M. 2008,
  ApJL, 685, L67, \dodoi{10.1086/592401}

\bibitem[{{Deller} {et~al.}(2019){Deller}, {Goss}, {Brisken}, {Chatterjee},
  {Cordes}, {Janssen}, {Kovalev}, {Lazio}, {Petrov}, {Stappers}, \&
  {Lyne}}]{2019ApJ...875..100D}
{Deller}, A.~T., {Goss}, W.~M., {Brisken}, W.~F., {et~al.} 2019, ApJ, 875, 100,
  \dodoi{10.3847/1538-4357/ab11c7}

\bibitem[{{Dolch} {et~al.}(2016){Dolch}, {Chatterjee}, {Clemens}, {Cordes},
  {Cashmen}, \& {Taylor}}]{2016JASS...33..167D}
{Dolch}, T., {Chatterjee}, S., {Clemens}, D.~P., {et~al.} 2016, Journal of
  Astronomy and Space Sciences, 33, 167, \dodoi{10.5140/JASS.2016.33.3.167}

\bibitem[{{Draine}(2011)}]{2011piim.book.....D}
{Draine}, B.~T. 2011, {Physics of the Interstellar and Intergalactic Medium}

\bibitem[{{Feng} {et~al.}(2020){Feng}, {Li}, {Long}, {Bellazzini}, {Costa},
  {Wu}, {Huang}, {Jiang}, {Minuti}, {Wang}, {Xu}, {Yang}, {Baldini}, {Citraro},
  {Nasimi}, {Soffitta}, {Muleri}, {Jung}, {Yu}, {Jin}, {Zeng}, {An}, {Brez},
  {Latronico}, {Sgro}, {Spandre}, \& {Pinchera}}]{2020NatAs...4..511F}
{Feng}, H., {Li}, H., {Long}, X., {et~al.} 2020, Nature Astronomy, 4, 511,
  \dodoi{10.1038/s41550-020-1088-1}

\bibitem[{{Gaensler} {et~al.}(2002){Gaensler}, {Jones}, \&
  {Stappers}}]{2002ApJ...580L.137G}
{Gaensler}, B.~M., {Jones}, D.~H., \& {Stappers}, B.~W. 2002, ApJL, 580, L137,
  \dodoi{10.1086/345750}

\bibitem[{{Gautam} {et~al.}(2013){Gautam}, {Chatterjee}, {Cordes}, {Deller}, \&
  {LAZIO}}]{2013AAS...22115404G}
{Gautam}, A., {Chatterjee}, S., {Cordes}, J.~M., {Deller}, A.~T., \& {LAZIO},
  J. 2013, in American Astronomical Society Meeting Abstracts, Vol. 221,
  American Astronomical Society Meeting Abstracts \#221, 154.04

\bibitem[{{Gurnett} \& {Kurth}(2019)}]{2019NatAs...3.1024G}
{Gurnett}, D.~A., \& {Kurth}, W.~S. 2019, Nature Astronomy, 3, 1024,
  \dodoi{10.1038/s41550-019-0918-5}

\bibitem[{{Gurnett} {et~al.}(2013){Gurnett}, {Kurth}, {Burlaga}, \&
  {Ness}}]{2013Sci...341.1489G}
{Gurnett}, D.~A., {Kurth}, W.~S., {Burlaga}, L.~F., \& {Ness}, N.~F. 2013,
  Science, 341, 1489, \dodoi{10.1126/science.1241681}

\bibitem[{{Gurnett} {et~al.}(2015){Gurnett}, {Kurth}, {Stone}, {Cummings},
  {Krimigis}, {Decker}, {Ness}, \& {Burlaga}}]{2015ApJ...809..121G}
{Gurnett}, D.~A., {Kurth}, W.~S., {Stone}, E.~C., {et~al.} 2015, ApJ, 809, 121,
  \dodoi{10.1088/0004-637X/809/2/121}

\bibitem[{{Gurnett} {et~al.}(2021){Gurnett}, {Kurth}, {Stone}, {Cummings},
  {Heikkila}, {Lal}, {Krimigis}, {Decker}, {Ness}, \&
  {Burlaga}}]{2021AJ....161...11G}
---. 2021, AJ, 161, 11, \dodoi{10.3847/1538-3881/abc337}

\bibitem[{{Jankowski} {et~al.}(2021){Jankowski}, {Keane}, \&
  {Stappers}}]{2021MNRAS.504..406J}
{Jankowski}, F., {Keane}, E.~F., \& {Stappers}, B.~W. 2021, MNRAS, 504, 406,
  \dodoi{10.1093/mnras/stab824}

\bibitem[{{Janssen} \& {Stappers}(2006)}]{2006AA...457..611J}
{Janssen}, G.~H., \& {Stappers}, B.~W. 2006, AAP, 457, 611,
  \dodoi{10.1051/0004-6361:20065267}

\bibitem[{{Jones} {et~al.}(2002){Jones}, {Stappers}, \&
  {Gaensler}}]{2002AA...389L...1J}
{Jones}, D.~H., {Stappers}, B.~W., \& {Gaensler}, B.~M. 2002, A\&A, 389, L1,
  \dodoi{10.1051/0004-6361:20020651}

\bibitem[{{Jones} {et~al.}(2017){Jones}, {McLaughlin}, {Lam}, {Cordes},
  {Levin}, {Chatterjee}, {Arzoumanian}, {Crowter}, {Demorest}, {Dolch},
  {Ellis}, {Ferdman}, {Fonseca}, {Gonzalez}, {Jones}, {Lazio}, {Nice},
  {Pennucci}, {Ransom}, {Stinebring}, {Stairs}, {Stovall}, {Swiggum}, \&
  {Zhu}}]{2017ApJ...841..125J}
{Jones}, M.~L., {McLaughlin}, M.~A., {Lam}, M.~T., {et~al.} 2017, ApJ, 841,
  125, \dodoi{10.3847/1538-4357/aa73df}

\bibitem[{{Kargaltsev} {et~al.}(2017){Kargaltsev}, {Pavlov}, {Klingler}, \&
  {Rangelov}}]{2017JPlPh..83e6301K}
{Kargaltsev}, O., {Pavlov}, G.~G., {Klingler}, N., \& {Rangelov}, B. 2017,
  Journal of Plasma Physics, 83, 635830501, \dodoi{10.1017/S0022377817000630}

\bibitem[{{Kramer} {et~al.}(2006){Kramer}, {Lyne}, {O'Brien}, {Jordan}, \&
  {Lorimer}}]{2006Sci...312..549K}
{Kramer}, M., {Lyne}, A.~G., {O'Brien}, J.~T., {Jordan}, C.~A., \& {Lorimer},
  D.~R. 2006, Science, 312, 549, \dodoi{10.1126/science.1124060}

\bibitem[{{Krishnakumar} {et~al.}(2015){Krishnakumar}, {Mitra}, {Naidu},
  {Joshi}, \& {Manoharan}}]{2015ApJ...804...23K}
{Krishnakumar}, M.~A., {Mitra}, D., {Naidu}, A., {Joshi}, B.~C., \&
  {Manoharan}, P.~K. 2015, ApJ, 804, 23, \dodoi{10.1088/0004-637X/804/1/23}

\bibitem[{{Lam} {et~al.}(2016){Lam}, {Cordes}, {Chatterjee}, {Jones},
  {McLaughlin}, \& {Armstrong}}]{2016ApJ...821...66L}
{Lam}, M.~T., {Cordes}, J.~M., {Chatterjee}, S., {et~al.} 2016, ApJ, 821, 66,
  \dodoi{10.3847/0004-637X/821/1/66}

\bibitem[{{Lee} \& {Lee}(2019)}]{2019NatAs...3..154L}
{Lee}, K.~H., \& {Lee}, L.~C. 2019, Nature Astronomy, 3, 154,
  \dodoi{10.1038/s41550-018-0650-6}

\bibitem[{{Lee} \& {Lee}(2020)}]{2020ApJ...904...66L}
---. 2020, ApJ, 904, 66, \dodoi{10.3847/1538-4357/abba20}

\bibitem[{{Linsky} {et~al.}(2019){Linsky}, {Redfield}, \&
  {Tilipman}}]{2019ApJ...886...41L}
{Linsky}, J.~L., {Redfield}, S., \& {Tilipman}, D. 2019, ApJ, 886, 41,
  \dodoi{10.3847/1538-4357/ab498a}

\bibitem[{{Liu} {et~al.}(2021){Liu}, {Wang}, {Yan}, {Shen}, {Tong}, {Huang}, \&
  {Zhao}}]{2021ApJ...912...58L}
{Liu}, J., {Wang}, H.-G., {Yan}, Z., {et~al.} 2021, ApJ, 912, 58,
  \dodoi{10.3847/1538-4357/abf140}

\bibitem[{{Manchester} {et~al.}(2005){Manchester}, {Hobbs}, {Teoh}, \&
  {Hobbs}}]{2005AJ....129.1993M}
{Manchester}, R.~N., {Hobbs}, G.~B., {Teoh}, A., \& {Hobbs}, M. 2005, AJ, 129,
  1993, \dodoi{10.1086/428488}

\bibitem[{{McComas} {et~al.}(2012){McComas}, {Alexashov}, {Bzowski}, {Fahr},
  {Heerikhuisen}, {Izmodenov}, {Lee}, {M{\"o}bius}, {Pogorelov}, {Schwadron},
  \& {Zank}}]{2012Sci...336.1291M}
{McComas}, D.~J., {Alexashov}, D., {Bzowski}, M., {et~al.} 2012, Science, 336,
  1291, \dodoi{10.1126/science.1221054}

\bibitem[{{McComas} {et~al.}(2015){McComas}, {Bzowski}, {Frisch}, {Fuselier},
  {Kubiak}, {Kucharek}, {Leonard}, {M{\"o}bius}, {Schwadron}, {Sok{\'o}{\l}},
  {Swaczyna}, \& {Witte}}]{2015ApJ...801...28M}
{McComas}, D.~J., {Bzowski}, M., {Frisch}, P., {et~al.} 2015, ApJ, 801, 28,
  \dodoi{10.1088/0004-637X/801/1/28}

\bibitem[{{Mignani} {et~al.}(2016){Mignani}, {Testa}, {Marelli}, {De Luca},
  {Salvetti}, {Belfiore}, {Pierbattista}, {Razzano}, {Shearer}, \&
  {Moran}}]{2016ApJ...825..151M}
{Mignani}, R.~P., {Testa}, V., {Marelli}, M., {et~al.} 2016, ApJ, 825, 151,
  \dodoi{10.3847/0004-637X/825/2/151}

\bibitem[{{Ocker} {et~al.}(2021){Ocker}, {Cordes}, {Chatterjee}, {Gurnett},
  {Kurth}, \& {Spangler}}]{2021NatAs.tmp...84O}
{Ocker}, S.~K., {Cordes}, J.~M., {Chatterjee}, S., {et~al.} 2021, Nature
  Astronomy, \dodoi{10.1038/s41550-021-01363-7}

\bibitem[{{Opher}(2016)}]{2016SSRv..200..475O}
{Opher}, M. 2016, Space Sci. Rev., 200, 475, \dodoi{10.1007/s11214-015-0186-3}

\bibitem[{{Palfreyman} {et~al.}(2018){Palfreyman}, {Dickey}, {Hotan},
  {Ellingsen}, \& {van Straten}}]{2018Natur.556..219P}
{Palfreyman}, J., {Dickey}, J.~M., {Hotan}, A., {Ellingsen}, S., \& {van
  Straten}, W. 2018, Nature, 556, 219, \dodoi{10.1038/s41586-018-0001-x}

\bibitem[{{Peri} {et~al.}(2012){Peri}, {Benaglia}, {Brookes}, {Stevens}, \&
  {Isequilla}}]{2012A&A...538A.108P}
{Peri}, C.~S., {Benaglia}, P., {Brookes}, D.~P., {Stevens}, I.~R., \&
  {Isequilla}, N.~L. 2012, A\&A, 538, A108, \dodoi{10.1051/0004-6361/201118116}

\bibitem[{{Peri} {et~al.}(2015){Peri}, {Benaglia}, \&
  {Isequilla}}]{2015A&A...578A..45P}
{Peri}, C.~S., {Benaglia}, P., \& {Isequilla}, N.~L. 2015, A\&A, 578, A45,
  \dodoi{10.1051/0004-6361/201424676}

\bibitem[{{Pletsch} {et~al.}(2012){Pletsch}, {Guillemot}, {Allen}, {Kramer},
  {Aulbert}, {Fehrmann}, {Ray}, {Barr}, {Belfiore}, {Camilo}, {Caraveo},
  {{\c{C}}elik}, {Champion}, {Dormody}, {Eatough}, {Ferrara}, {Freire},
  {Hessels}, {Keith}, {Kerr}, {de Luca}, {Lyne}, {Marelli}, {McLaughlin},
  {Parent}, {Ransom}, {Razzano}, {Reich}, {Saz Parkinson}, {Stappers}, \&
  {Wolff}}]{2012ApJ...744..105P}
{Pletsch}, H.~J., {Guillemot}, L., {Allen}, B., {et~al.} 2012, ApJ, 744, 105,
  \dodoi{10.1088/0004-637X/744/2/105}

\bibitem[{{Rangelov} {et~al.}(2016){Rangelov}, {Pavlov}, {Kargaltsev},
  {Durant}, {Bykov}, \& {Krassilchtchikov}}]{2016ApJ...831..129R}
{Rangelov}, B., {Pavlov}, G.~G., {Kargaltsev}, O., {et~al.} 2016, ApJ, 831,
  129, \dodoi{10.3847/0004-637X/831/2/129}

\bibitem[{{Rangelov} {et~al.}(2017){Rangelov}, {Pavlov}, {Kargaltsev},
  {Reisenegger}, {Guillot}, {van Kerkwijk}, \& {Reyes}}]{2017ApJ...835..264R}
---. 2017, ApJ, 835, 264, \dodoi{10.3847/1538-4357/835/2/264}

\bibitem[{{Reardon} {et~al.}(2016){Reardon}, {Hobbs}, {Coles}, {Levin},
  {Keith}, {Bailes}, {Bhat}, {Burke-Spolaor}, {Dai}, {Kerr}, {Lasky},
  {Manchester}, {Os{\l}owski}, {Ravi}, {Shannon}, {van Straten}, {Toomey},
  {Wang}, {Wen}, {You}, \& {Zhu}}]{2016MNRAS.455.1751R}
{Reardon}, D.~J., {Hobbs}, G., {Coles}, W., {et~al.} 2016, MNRAS, 455, 1751,
  \dodoi{10.1093/mnras/stv2395}

\bibitem[{{Redfield} \& {Linsky}(2004)}]{2004ApJ...613.1004R}
{Redfield}, S., \& {Linsky}, J.~L. 2004, ApJ, 613, 1004, \dodoi{10.1086/423311}

\bibitem[{{Romani} {et~al.}(2010){Romani}, {Shaw}, {Camilo}, {Cotter}, \&
  {Sivakoff}}]{2010ApJ...724..908R}
{Romani}, R.~W., {Shaw}, M.~S., {Camilo}, F., {Cotter}, G., \& {Sivakoff},
  G.~R. 2010, ApJ, 724, 908, \dodoi{10.1088/0004-637X/724/2/908}

\bibitem[{{Romani} {et~al.}(2017){Romani}, {Slane}, \&
  {Green}}]{2017ApJ...851...61R}
{Romani}, R.~W., {Slane}, P., \& {Green}, A.~W. 2017, ApJ, 851, 61,
  \dodoi{10.3847/1538-4357/aa9890}

\bibitem[{{Shaw} {et~al.}(2018){Shaw}, {Lyne}, {Stappers}, {Weltevrede},
  {Bassa}, {Lien}, {Mickaliger}, {Breton}, {Jordan}, {Keith}, \&
  {Krimm}}]{2018MNRAS.478.3832S}
{Shaw}, B., {Lyne}, A.~G., {Stappers}, B.~W., {et~al.} 2018, MNRAS, 478, 3832,
  \dodoi{10.1093/mnras/sty1294}

\bibitem[{{Shemar} \& {Lyne}(1996)}]{1996MNRAS.282..677S}
{Shemar}, S.~L., \& {Lyne}, A.~G. 1996, MNRAS, 282, 677,
  \dodoi{10.1093/mnras/282.2.677}

\bibitem[{{Shklovskii}(1970)}]{1970SvA....13..562S}
{Shklovskii}, I.~S. 1970, SvA, 13, 562

\bibitem[{{Swaczyna} {et~al.}(2018){Swaczyna}, {Bzowski}, {Kubiak},
  {Sok{\'o}{\l}}, {Fuselier}, {Galli}, {Heirtzler}, {Kucharek}, {McComas},
  {M{\"o}bius}, {Schwadron}, \& {Wurz}}]{2018ApJ...854..119S}
{Swaczyna}, P., {Bzowski}, M., {Kubiak}, M.~A., {et~al.} 2018, ApJ, 854, 119,
  \dodoi{10.3847/1538-4357/aaabbf}

\bibitem[{{Toropina} {et~al.}(2019){Toropina}, {Romanova}, \&
  {Lovelace}}]{2019MNRAS.484.1475T}
{Toropina}, O.~D., {Romanova}, M.~M., \& {Lovelace}, R.~V.~E. 2019, MNRAS, 484,
  1475, \dodoi{10.1093/mnras/stz034}

\bibitem[{{Vigelius} {et~al.}(2007){Vigelius}, {Melatos}, {Chatterjee},
  {Gaensler}, \& {Ghavamian}}]{2007MNRAS.374..793V}
{Vigelius}, M., {Melatos}, A., {Chatterjee}, S., {Gaensler}, B.~M., \&
  {Ghavamian}, P. 2007, MNRAS, 374, 793,
  \dodoi{10.1111/j.1365-2966.2006.11193.x}

\bibitem[{{Wilkin}(1996)}]{1996ApJ...459L..31W}
{Wilkin}, F.~P. 1996, ApJL, 459, L31, \dodoi{10.1086/309939}

\bibitem[{{Yoon} \& {Heinz}(2017)}]{2017MNRAS.464.3297Y}
{Yoon}, D., \& {Heinz}, S. 2017, MNRAS, 464, 3297,
  \dodoi{10.1093/mnras/stw2590}

\bibitem[{{Yuan} {et~al.}(2010){Yuan}, {Wang}, {Manchester}, \&
  {Liu}}]{2010MNRAS.404..289Y}
{Yuan}, J.~P., {Wang}, N., {Manchester}, R.~N., \& {Liu}, Z.~Y. 2010, MNRAS,
  404, 289, \dodoi{10.1111/j.1365-2966.2010.16272.x}

\bibitem[{{Zank} {et~al.}(2013){Zank}, {Heerikhuisen}, {Wood}, {Pogorelov},
  {Zirnstein}, \& {McComas}}]{2013ApJ...763...20Z}
{Zank}, G.~P., {Heerikhuisen}, J., {Wood}, B.~E., {et~al.} 2013, ApJ, 763, 20,
  \dodoi{10.1088/0004-637X/763/1/20}

\bibitem[{{Zank} {et~al.}(2019){Zank}, {Nakanotani}, \&
  {Webb}}]{2019ApJ...887..116Z}
{Zank}, G.~P., {Nakanotani}, M., \& {Webb}, G.~M. 2019, ApJ, 887, 116,
  \dodoi{10.3847/1538-4357/ab528c}

\bibitem[{{Zieger} {et~al.}(2013){Zieger}, {Opher}, {Schwadron}, {McComas}, \&
  {T{\'o}th}}]{2013GeoRL..40.2923Z}
{Zieger}, B., {Opher}, M., {Schwadron}, N.~A., {McComas}, D.~J., \& {T{\'o}th},
  G. 2013, GRL, 40, 2923, \dodoi{10.1002/grl.50576}

\end{thebibliography}

\listofchanges

\end{document}